\documentclass{ar3e} %for draft 
\begin{document}
\input epsf.tex    %<-If you need EPS figures to be
                   %  called in {figure} environment for PC
%\input epsf.def   %<-If you need EPS figures to be
                   %  called in {figure} environment for Macintosh

% Definitions
\def\nuc#1#2{${}^{#1}$#2}
\def\mee{$\langle m_{ee} \rangle$}
\def\mnu{$\langle m_{\nu} \rangle$}
\def\gnu{$\langle g_{\nu,\chi}\rangle$}
\def\mmod{$\| \langle m_{ee} \rangle \|$}
\def\mb{$\langle m_{\beta} \rangle$}
\def\BBz{$\beta\beta(0\nu)$}
\def\BBm{$\beta\beta(0\nu,\chi)$}
\def\BBt{$\beta\beta(2\nu)$}
\def\BB{$\beta\beta$}
\def\Mz{$|M_{0\nu}|$}
\def\Mt{$|M_{2\nu}|$}
\def\Tz{$T^{0\nu}_{1/2}$}
\def\Tt{$T^{2\nu}_{1/2}$}
\def\Tc{$T^{0\nu\,\chi}_{1/2}$}
\def\be{\begin{equation}}
\def\ee{\end{equation}}
\def\today{\space\number\day\space\ifcase\month\or January\or February\or
  March\or April\or May\or June\or July\or August\or September\or October\or
  November\or December\fi\space\number\year}
% end definitions

\input psfig.sty

%\jname{Annu. Rev. Nucl. Part. Sci.}
\jname{Submitted to Annu. Rev. Nucl. Part. Sci.} % for preprint
%\jyear{2002}
%\jvol{52}
\jvol{(2002)} % for preprint
\jyear{{\bf 52} }  % for preprint
%\ARinfo{1056-8700/97/0610-00}

\title{Double Beta Decay}

\markboth{Elliott and Vogel}{Double Beta Decay}

\author{Steven R. Elliott 
\affiliation{Department of Physics, University of Washington, Seattle, Washington 98195}
Petr Vogel 
\affiliation{Department of Physics, California Institute of Technology, Pasadena, California 91125}}


\begin{abstract}
The motivation, present status, and future plans
of the search for \BBz\ decay are reviewed. It is argued
that, motivated by the recent observations of neutrino oscillations,
there is a reasonable hope that \BBz\ decay corresponding
to the neutrino mass scale suggested by oscillations,
$m_{\nu} \approx$ 50 meV, actually exists. The challenges
to achieve the sensitivity corresponding to this
mass scale, and plans to overcome them, are described.
\end{abstract}

\maketitle

\section{INTRODUCTION}
%-----------------------------
%
%	last update: 2/17/02
%
%-----------------------------

Since the last Annual Review article on double beta decay \cite{MV94}, published in 1994,
there have been several exciting developments. Most significantly, the neutrino
oscillation experiments convincingly
show that neutrinos have a finite mass. However, in oscillation
experiments only the differences in squares of the neutrino masses, 
$\Delta m^{2}_{ij} \equiv | m_i^2 - m_j^2| $, can be measured.
Nevertheless, a lower limit on the absolute value  of the neutrino mass scale, 
$m_{scale} =  \sqrt{\Delta m^2}$, has been
established in this way. Its existence, in turn, is causing a renaissance
of enthusiasm in the double beta decay community which is expected to reach, in
the next generation of experiments, the sensitivity corresponding to this mass scale. 
Below we review the current status of the double beta decay and the effort devoted
to reach the required sensitivity. But before proceeding, we briefly summarize
the achievements of the neutrino oscillation searches and the role that the search
for the neutrinoless double beta decay plays in the elucidation of the pattern
of neutrino masses and mixing. In these introductory remarks we use the 
established terminology, some of which will only be defined later in the text.

There is a consensus that the measurement of atmospheric
neutrinos by the SuperKamiokande collaboration \cite{SKatm01}
can be only interpreted as a consequence
of the nearly maximum mixing between  $\nu_{\mu}$ and $\nu_{\tau}$ neutrinos
(i.e., $\sin^2 2\theta_{atm} \sim 1.0$) 
with the corresponding mass squared difference  
$\Delta m_{atm}^2 \sim 3\times10^{-3}{\rm eV}^2$,
i.e., $m_{scale} \sim$ 50 meV.
This finding is supported  by the K2K experiment
\cite{K2K01} that uses an accelerator  
$\nu_{\mu}$ beam pointing towards the SuperKamiokande detector
250 km away,
and is in accord with the earlier findings of the Kamiokande \cite{Kam94},
IMB \cite{IMB92} and Soudan \cite{SOU99} experiments.
 Several large long-baseline experiments are being built to further elucidate
this discovery, and determine the corresponding parameters more accurately. 

At the same time the `solar neutrino puzzle', which has been with us for over thirty years
since the pioneering chlorine experiment of Davis \cite{chlorine}, also reached
the stage where the interpretation of the measurements in terms of  oscillations
between the $\nu_e$ and some combination of active, 
i.e. $\nu_{\mu}$ and $\nu_{\tau}$ neutrinos, seems inescapable. In particular, the
juxtaposition of the results of the
SNO experiment \cite{SNO01} and SuperKamiokande \cite{SKsol01},
together with the earlier solar neutrino flux determination in
the chlorine and gallium \cite{Gallex,Sage} experiments,
leads to that conclusion.
The value of the corresponding oscillation parameters remain uncertain, with several
`solutions' possible, although the so-called Large Mixing Angle (LMA) solution
with $\sin^2 2\theta_{sol} \sim 0.8$ (but  $\sin^2 2\theta_{sol} < 1)$ and 
$\Delta m_{sol}^2 \sim 5\times10^{-5}{\rm eV}^2$ is preferred at present. 
Again, the continuing
and soon to be operational experiments, like KamLAND and Borexino, aim
to find with more certainty which of the possible solutions is the correct one.

The pattern of neutrino mixing is further simplified by the constraint due to
the \textsc{Chooz} and \textsc{Palo Verde} 
reactor neutrino experiments \cite{Chooz99,Palo01}
which lead to the
conclusion that the third mixing angle, $\theta_{13}$, is
small,  $\sin^2 2\theta_{13} \le 0.1$.  

The oscillation experiments cannot determine the absolute
magnitude of the masses and, in particular, cannot at this stage separate
two rather different scenarios, the hierarchical pattern of neutrino
masses in which the neutrino masses $m_i$ and/or $m_j$ are of similar
magnitude as 
$ \sqrt{\Delta m_{ij}^2}$ and the degenerate pattern
in which  all $m_i \gg \sqrt{\Delta m_{ij}^2}$. It is likely that the 
search for the neutrinoless double beta
decay, reviewed here, will help in the foreseeable future
in establishing the correct mass pattern
and in determining, or at least strongly constraining, the absolute neutrino mass
scale.

Moreover, the oscillation results do not tell us anything about the properties of neutrinos
under charge conjugation. While the charged leptons are Dirac particles, distinct from
their antiparticles, neutrinos may be the ultimate neutral particles, as
envisioned by Majorana, identical with their antiparticles. That fundamental
distinction becomes important only for massive particles.
Neutrinoless double beta decay proceeds only when neutrinos are massive 
Majorana particles, hence its observation would resolve the question.

Quite generally, the Standard Electroweak Model postulates that neutrinos are massless,
and that the total lepton number, as well as the individual flavor lepton numbers,
are conserved. For various reasons most people believe that the model, despite 
its enormous success, is incomplete, and the hunt for `physics beyond the Standard
Model' is being actively pursued on many fronts. Observation of neutrino mass
and oscillation is clearly an example of a phenomenon that is at variance with the
Standard Model. Further elucidation of various aspects of neutrino mass 
will undoubtedly help
in pointing to the proper generalization of the Standard Model. 

Double beta decay is a rare transition between two nuclei with the
same mass number $A$ involving change of the nuclear charge $Z$
by two units. The decay can proceed only if the initial nucleus is
less bound than the final one, and both must be more bound than
the intermediate nucleus
(or the decay to the intermediate nucleus must be highly hindered,
as in $^{48}$Ca). These conditions are fulfilled in nature
for many even-even nuclei, and only for them. Typically, the decay 
can proceed from the ground state (spin and parity always $0^+$) of the
initial nucleus to the ground state (also  $0^+$) of the final
nucleus, although the decay into  excited states
($0^+$ or $2^+$) is in some cases also energetically possible.

The two-neutrino decay, \BBt,
\be
(Z,A)  \rightarrow (Z+2,A) + e_1^- + e_2^- + \bar{\nu}_{e1}  + \bar{\nu}_{e2}
\label{e:2nu}
\ee
conserves not only electric charge but also
lepton number. 
(Analogous decays, involving transformation of two protons into two
neutrons, are also sometimes possible. We concentrate here, however,
on the decays $2n \rightarrow 2p$ with more candidate nuclei,
and usually larger $Q$ values.)

On the other hand,
the neutrinoless decay, \BBz,
\be
(Z,A)  \rightarrow (Z+2,A) + e_1^- + e_2^-
\label{e:0nu}
\end{equation}
violates lepton number conservation and is therefore
forbidden in the standard
 electroweak theory. In addition, there can be transitions
\BBm\ in which a light neutral
boson $\chi$, a Majoron postulated in various extensions of
the standard electroweak theory \cite{Maj81},
is emitted:
\begin{equation}
(Z,A)  \rightarrow (Z+2,A) + e_1^- + e_2^-  + \chi ~.
\label{e:mnu}
\end{equation}

It is fascinating to realize that the interest in \BB\ decay
spans more than six decades. Already in 1937 Racah \cite{Rac37},
following the fundamental suggestion of Majorana \cite{Maj37},
discussed the possibility of a neutrinoless transformation
of two neutrons into two protons plus two electrons. Even earlier,
Goeppert-Mayer \cite{May39} evaluated the
decay rate of the \BBt\ mode and realized that the corresponding 
half-life could exceed $10^{20}$ y. And Furry, shortly afterwards \cite{Fur39},
estimated that the \BBz\ decay should be much faster than the \BBt\
decay. That conclusion, however, had to be revised with the
discovery of parity nonconservation in weak interactions.
Thus, the stage was set for the realization that the
observation of the \BBz\ decay would establish that
the neutrino is a massive Majorana particle.

There were numerous earlier reviews of \BB\ decay, including the
classics by Primakoff \& Rosen \cite{PR59}, Haxton \& Stephenson
\cite{HS84}, and Doi, Kotani \& Takasugi \cite{DKT85}. 
 More recent reviews,
besides \cite{MV94}, include those by Boehm \& Vogel \cite{BV92}
(where the phase-space integrals are listed),
Suhonen \& Civitarese \cite{SC98}
(which deals mostly with the nuclear matrix elements), Faessler \& \v{S}imkovic
\cite{FS98}, Vergados \cite{Ver00}, and Klapdor-Kleingrothaus \cite{Kla00}.
A rather complete list of experimental data can be found in the papers
by Tretyak and Zdesenko \cite{TZ95,TZ02}.  
In the Review of Particle Physics \cite{RPP} the most important $\beta\beta$
decay experimental data are regularly listed. The whole field of neutrino
mass and oscillations has been reviewed recently in Annual Reviews
\cite{FKM99}. 

\begin{figure}[htb]%1
%\centerline{\epsfbox{figure1.eps}}
\epsfysize=4.5in \epsfbox{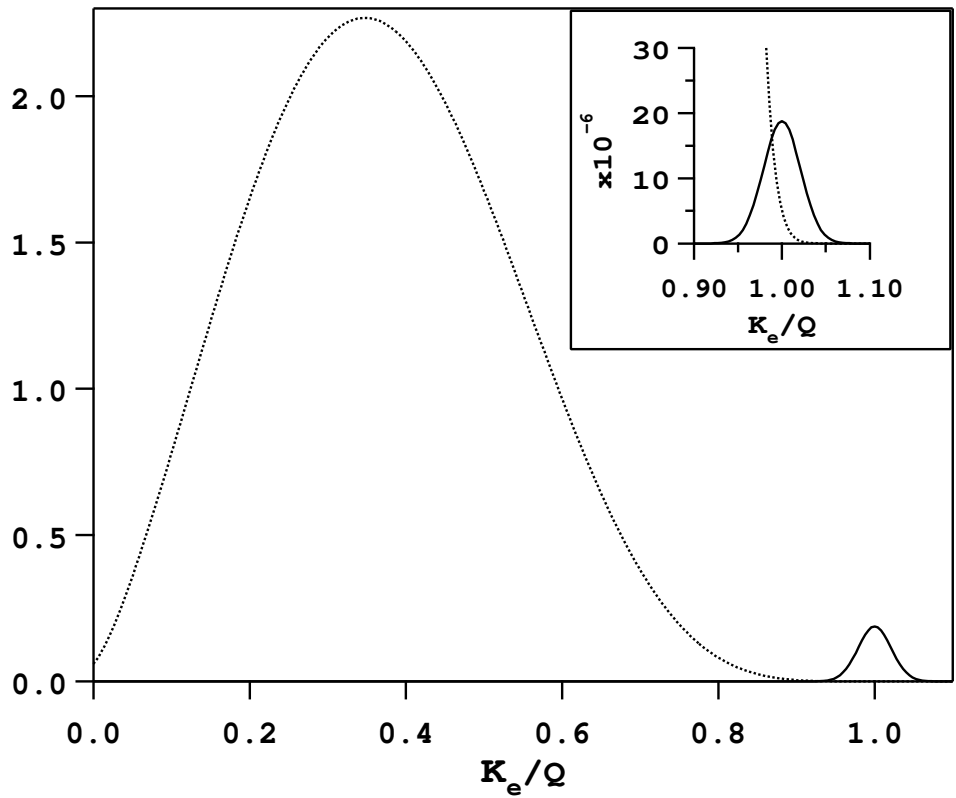} 
\caption{Illustration of the spectra of the sum of the electron kinetic energies 
K$_e$ (Q is the endpoint) for the 
\BBt\, normalized to 1 (dotted curve)
and \BBz\ decays (solid curve). The \BBz\ spectrum is normalized to $10^{-2}$ 
($10^{-6}$ in the figure inset).
All spectra are convolved with
an energy resolution of 5\%, representative of several
experiments.  However, some experiments, notably Ge, have  
a much better energy resolution. }
\label{fig:spect}
\end{figure}

It is easy to distinguish the three decay modes (\BBt, \BBz, and \BBm)
by the shape of the electron sum energy spectra, which 
are determined by the phase space of the outgoing light particles.
For the \BBt\ and \BBz\ modes these spectra
are illustrated in Figure \ref{fig:spect}.
In the $2\nu$ decay there is a broad maximum at the sum
kinetic energy of the two electrons below half of
the endpoint energy.  
In contrast, in the \BBz\ mode the two electrons carry the full available kinetic
energy  (the nuclear recoil is
negligible for all practical purposes)
and the spectrum is therefore a single peak at the endpoint. 
In the Majoron
decay, not shown in order not to clutter the figure, 
the electron spectrum is again continuous, but
the maximum is shifted higher, above the halfway point,
as required by the three-body light particle phase
space.

The insert in Fig. \ref{fig:spect} illustrates in detail
the expected spectra near the endpoint where the \BBt\ decay
represents the ultimate background in the search of the  
 \BBz\ mode. (See section 4.2.4 for the discussion of this
point.)

The \BBt\ decay mode is an allowed process.
However, since it is a second order semileptonic weak decay,
its lifetime, proportional to $(G_F \cos \theta_C)^{-4}$,
is very long. The observation of the \BBt\ decay  
presents a formidable challenge,
since it must be detected despite the presence of inevitable
traces of radioisotopes with similar decay energy, but
lifetimes more than 10 orders of magnitude shorter.
Nevertheless, at present, that challenge has been met and
the \BBt\ decay  has been positively identified
in a number of cases.
Observing the \BBt\ decay is important not only as a proof
that the necessary background suppression has been achieved,
but also allows one to constrain the nuclear models needed
to evaluate the corresponding nuclear matrix elements.

\begin{figure}[htb]
%\epsfscale1200         % Figure enlarged to 120 (MAC)%
%\epsfxsize10pc         %
\epsfysize=6.5in \epsfbox{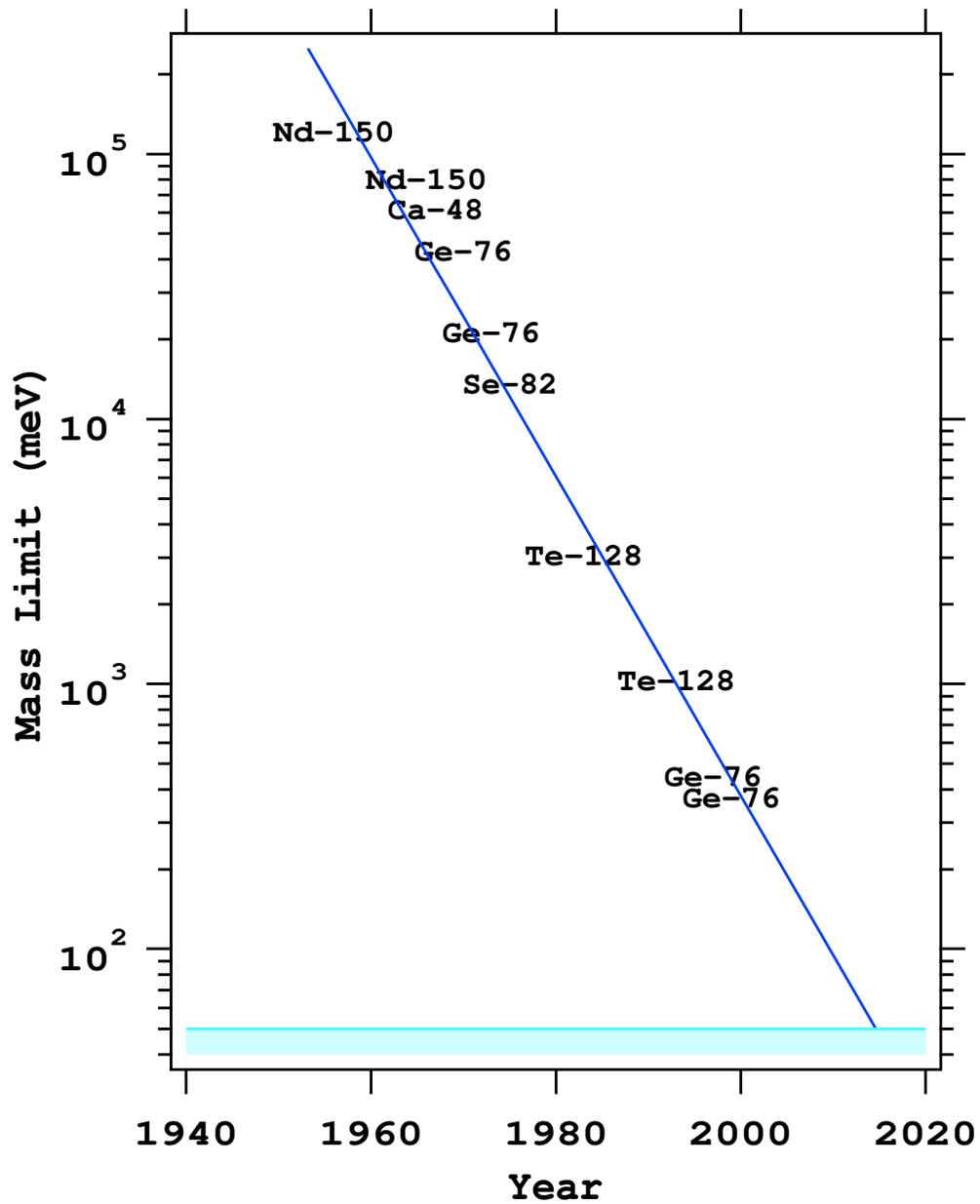}
%\centerline{\epsfbox{HistoryGraph3.EPSF}}
\caption{``Moore's law'' of \BBz\ decay: the limit
of the effective neutrino mass vs. time. The corresponding experiments
are denoted by the symbol for the initial nucleus. The uncertainty
in the nuclear matrix elements is not included in this illustration.
The gray band near the bottom indicates the neutrino mass scale
$\protect\sqrt{\Delta m_{atm}^2}\protect\ $ .
} 
\label{fig:history}
\end{figure}

Hand in hand with the observation of the \BBt\ decays, the 
experiments became sensitive to longer and longer half-lives
of the \BBz\ decay mode. Since the rate of  \BBz\ is proportional to
the square of effective neutrino mass, the improvements lead
to correspondingly improved limits on the mass. 
This is illustrated in
Figure \ref{fig:history}, which shows an essentially
logarithmic improvement, by a factor larger than about four
every decade, of the corresponding limits. If this trend
continues, we expect to reach the neutrino mass scale  suggested by
the oscillation experiments in 10-20 years. Given the
typical lead time of the large particle physics experiments,
it suggests that the relevant \BB\ decay experiments 
should enter the `incubation' process now.

\section{NEUTRINO MASS - THEORETICAL ASPECTS}

%-----------------------------
%
%	last update: 2/17/02
%
%-----------------------------

\subsection{Majorana and Dirac neutrinos}

Empirically, neutrino masses are much smaller than the
masses of the charged leptons with which they form
weak isodoublets. 
Even the mass of the lightest charged lepton, the electron, is at least 
$10^5$ times larger than the neutrino mass constrained
by the tritium beta decay experiments. The existence of such large factors
is difficult to explain, unless one invokes some symmetry
principle. The assumption that neutrinos are 
Majorana particles is often used
in this context. Moreover, many theoretical constructs
invoked to explain neutrino masses lead to this conclusion.

The term `Majorana' is used for particles that are identical with their
own antiparticles while Dirac particles can be distinguished 
from their antiparticles.
This implies that Majorana fermions are two-component objects,
while Dirac fermions are four-component. 
In order to avoid confusion and to derive the
formula for the \BBz\ rate mediated by the exchange of massive Majorana
neutrinos, it is worthwhile to
discuss, albeit briefly, the formalism needed to describe them
(for more details, see e.g. \cite{Kobz80,DKT85,Kayser89,Zralek97}).

Massive
fermions are usually described by the
Dirac equation, where the chirality eigenstates
$\psi_R$ and $\psi_L$ are coupled and form
a four-component object of mass $m$,
\be
i(\hat{\sigma}^{\mu}\partial_{\mu})\psi_R - m\psi_L = 0~,~ 
i(\sigma^{\mu}\partial_{\mu})\psi_L - m\psi_R = 0 ~,   
\ee
where $\hat{\sigma}^{\mu} = (\sigma^0, \vec{\sigma})$,  
$\sigma^{\mu} = (\sigma^0, -\vec{\sigma})$ and $(\sigma^0, \vec{\sigma})$
are the Pauli matrices.
As written, $\psi_{L(R)}$ are two-component spinors; the usual
four-component bispinors are defined as:
\be
\Psi = \left( \begin{array}{c} 
\psi_R \\  \psi_L \end{array} \right) ~;~
\Psi_R = \left( \begin{array}{c} 
\psi_R \\  0      \end{array} \right) ~;~
\Psi_L = \left( \begin{array}{c} 
 0  \\   \psi_L      \end{array} \right) ~,
\label{eq:Dir_field}
\ee
where $\Psi_{L(R)}$ are just the chiral
projections of $\Psi$, i.e. the eigenstates of $P_{L(R)} = (1 \mp \gamma_5)/2$.

However, Majorana's suggestion \cite{Maj37} allows one to 
use an alternative description of those massive
fermions which do not have any additive quantum numbers as
either two-component
$\psi_R$ (mass $m$), or $\psi_L$ (mass $m'$), which
obey independent equations 
\be
i(\hat{\sigma}^{\mu}\partial_{\mu})\psi_R - m \epsilon \psi_R^* = 0 ~;~ 
i (\sigma^{\mu}\partial_{\mu})\psi_L + m' \epsilon \psi_L^* = 0 ~,
\ee
where $\epsilon = i\sigma_y$.

The Majorana fields can be also expressed in the four-component form
\be
\Psi_L(x) =  \left( \begin{array}{c} 
-\epsilon \psi_L^*(x) \\  \psi_L(x) \end{array} \right) ~,~
{\rm and/or} ~
\Psi_R(x) =  \left( \begin{array}{c}  \psi_R(x) \\
\epsilon  \psi_R^*(x) \end{array} \right) ~.
\label{eq:Maj_field}
\ee
Such a four-component notation is a convention useful to express
the charged weak current in a compact form. 
It is then clear that the Dirac field $\Psi$, Equation \ref{eq:Dir_field},
is equivalent to a pair of Majorana fields with $m = m'$ and
$ \psi_L = \epsilon  \psi_R^*$.

The four-component Majorana fields, Equation \ref{eq:Maj_field}, 
are selfconjugate, $\Psi_{L(R)}^c(x) = \Psi_{L(R)}(x)$, 
where charge conjugation is defined as
$\Psi_{L(R)}^c(x) = i\gamma^2\gamma^0 \bar{\Psi}_{L(R)}^T$. The fields  
$\Psi_L(x)$ and $\Psi_R(x)$ are eigenstates of $CP$ with opposite 
eigenvalues. 

The  Lorentz invariant mass term in  the neutrino Langrangian
can appear in three forms:
\be
M_D[\bar{\nu}_R \nu_L ~+~ (\bar{\nu}_L)^c  \nu_R^c] ~,~
M_L[(\bar{\nu}_L)^c \nu_L + \bar{\nu}_L \nu_L^c]  ~,~
M_R[(\bar{\nu}_R)^c \nu_R + \bar{\nu}_R \nu_R^c] ~,
\label{eq:massterm}
\ee
where we have introduced the notation $\nu_{L(R)}$ for the
corresponding neutrino annihilation operators.
The first expression in Equation \ref{eq:massterm} 
is the Dirac mass term (with the mass parameter $M_D$) which
requires the existence of both chirality eigenstates $\nu_L$ and $\nu_R$
and conserves the lepton quantum number. The second (and third)
mass terms are Majorana mass terms, which violate the lepton number
and can be present even without the existence of $\nu_R$ (for 
 the term with mass parameter $M_L$) or $\nu_L$ (for 
 the term with mass parameter $M_R$). In general, all three terms
might coexist, and then the mass Langrangian must be diagonalized
resulting in two generally nondegenerate mass eigenvalues for each flavor.
(That is the situation with the generic see-saw mass
\cite{Yan79}, where it
is assumed that
$M_R \gg M_D \gg M_L \sim 0$, and the light neutrino acquires the
mass $m_{\nu} \sim M_D^2/M_R$.) 

Let us consider now the general
situation with
$N$ flavors of the left-handed neutrinos $\nu_L$ and
in addition an equal number $N$ of the
right handed neutrinos $\nu_R$.
The most general Lorentz invariant mass term of the neutrino Langrangian
has then the form
\be
{\cal L}_M = -\frac{1}{2} \left( (\bar{\nu}_L)^c~~ \bar{\nu}_R \right)
{\cal M} \left( \begin{array}{c}
\nu_L \\ \nu_R^c   \end{array} \right) ~+{\rm h.c.} ~,~ 
{\cal M} = \left( \begin{array}{cc}
{\cal M}_L & {\cal M}_D^T \\  {\cal M}_D & {\cal M}_R \end{array} \right) ~,
\ee
where $\nu_L$ and $\nu_R$ are column vectors of dimension $N$.
Here ${\cal M}_L$ and  ${\cal M}_R$ are symmetric $N \times N$
matrices (Majorana masses for the
left- and right-handed neutrinos) and  ${\cal M}_D$ is an
arbitrary and generally complex $N \times N$ matrix.

The mass matrix ${\cal M}$,
with real positive eigenvalues  $m_1, \dots, m_{2N}$, 
is diagonalized by the $2N \times 2N$ unitary matrix
\be
\left( \begin{array}{c} \nu_L \\ \nu_R^c  \end{array} \right)
= \left( \begin{array}{c} U \\ V  \end{array} \right) \Phi_L ~.
\ee
The general mixing matrices $U$ and $V$ have $N$ rows and $2N$ columns
and $\Phi_L$ is a column vector of dimension $2N$ 
of Majorana-like objects \cite{Kobz80}.
On the other hand, if none of the states $\nu_R$ exist, 
or if ${\cal M}_R$ is so large that the corresponding
states need not be considered, only  ${\cal M}_L$
is relevant, and only the
$N \times N$ mixing matrix $U$ is needed to diagonalize
the mass term (and $\Phi_L$ has then only $N$ components, naturally).

Let us consider in more detail the latter case, 
when only  ${\cal M}_L \ne 0$.
The $N \times N$  unitary mixing matrix $U$ contains $N^2$ real parameters.
However, $N$ of them correspond to unphysical phases; there are
$N(N-1)/2$ angles and $N(N-1)/2$ physically relevant phases describing
possible CP violations. 
(For a discussion of parameter counting, see \cite{Kobz80}.)
In the oscillation experiments that violate only
the flavor lepton number, but conserve the total
lepton number (such as $\nu_e \rightarrow \nu_{\mu}$
or $\nu_{\mu} \rightarrow \nu_{\tau}$), one can determine,
in principle, all angles and $(N-1)(N-2)/2$ phases. These phases,
common to the Dirac and Majorana neutrinos, describe
$CP$ violation  responsible for the possible differences of the
oscillation probabilities $\nu_{\ell} \rightarrow \nu_{\ell'}$ and
$\bar{\nu}_{\ell} \rightarrow \bar{\nu}_{\ell'}$.

The remaining $N-1$ phases affect only neutrino
oscillation-like processes (in which neutrinos are
created in the charged current weak processes and absorbed
again in charged current) that violate the total
lepton number, such as the \BBz\ decay. Such phases are physically
significant only for Majorana neutrinos; they are unphysical
for Dirac neutrinos. This is so because for Majorana neutrinos
one cannot perform the transformation 
$\nu_i \rightarrow \nu_i' = e^{i\alpha_i}\nu_i$, which would
violate the selfconjugation property. 

In principle the distinction between Dirac and Majorana neutrinos
affects other processes as well, such as the angular distribution
of $\nu - e$ scattering or photon polarization in the 
$\nu_i \rightarrow \nu_j + \gamma$ decay. However,
the ``Practical Dirac-Majorana Confusion Theorem'' \cite{Kayser82}
states that the distinction vanishes for $m_{\nu} \rightarrow 0$
which makes it essentially unobservable in these cases.

\subsection {\BBz\ decay rate and Majorana mass}

Here we shall consider only the simplest case of
the left-handed $V - A$ weak currents
and light massive Majorana neutrinos. This is the case
of current interest provided the neutrino mass revealed in
the oscillation experiments is of Majorana character.
The more general expressions can be found e.g. in
the reviews \cite{HS84,DKT85}. (For recent formulation
of the general problem, see \cite{Pas99}.)

The differential decay rate of the \BBz\ process, Equation \ref{e:0nu},
is \cite{DKT85}
\be
d\Gamma_{0\nu} = 2\pi \sum_{spin} |R_{0\nu}|^2 
\delta({\epsilon_1 + \epsilon_2 + E_f - M_i}) 
\frac{{\rm d}\vec{p}_1}{(2\pi)^3} \frac{{\rm d}\vec{p}_2}{(2\pi)^3} ~,
\label{eq:0nurate}
\ee
where $\epsilon_{1(2)}$
and $\vec{p}_{1(2)}$ are total energies 
and momenta of the electrons and 
$E_f (M_i)$ is the
energy of the final (mass of the initial) nuclear state. The quantity
$R_{0\nu}$ is the reaction amplitude to be evaluated in the second
order pertubation theory with respect to the weak interactions. 

The lepton part of $R_{0\nu}$, involving the emission 
and reabsorption of the Majorana neutrino of mass $m_j$, is
\be
-i \int \frac{d^4 q}{(2 \pi )^4} e^{-iq(x-y)}
\bar{e}(x) \gamma_{\rho} P_L  
\frac{q^{\mu}\gamma_{\mu} + m_j}{q^2 - m_j^2} P_L 
\gamma_{\sigma} e^c(y) ~,
\label{eq:prop}
\ee
where $P_L = (1-\gamma_5)/2$,
$\bar{e}(x), e^c(y)$ are the electron
creation operators, and $q$ is the momentum transfer four-vector. 
Since $\gamma_{\mu}$ anticommute with $\gamma_5$,
this amplitude is proportional to 
$m_j$ and the term with $q^{\mu} \gamma_{\mu}$ vanishes. 
After integrating over the energy of the virtual neutrino $dq^0$,
the denominator $q^2 - m_j^2$ is replaced by its residue $\omega_j/\pi$,
where $\omega_j = \sqrt{\vec{q}^{~2} + m_j^2}$.
The amplitude is therefore proportional to $m_j/\omega_j \ll 1$ for light
neutrinos.
 
The remaining integration over the virtual neutrino momentum $\vec{q}$ leads
to the appearance of the neutrino potentials
\be
H_k (r, A_k) = \frac{2R_N}{\pi r} \int_0^{\infty} dq \frac{q \sin (qr)}
{\omega(\omega + A_k)} ~,~ A_{1(2)} = E_m - (M_i + M_f)/2 \pm 
(\epsilon_1 - \epsilon_2)/2 ~,
\ee
where 1 and 2 label the emitted electrons, 
$E_m$ is the excitation energy of the intermediate nucleus,
$M_f$ is the mass of the final nucleus,
and $r$ is the distance between the two neutrons that are
changed into protons. The factor $R_N$, the nuclear radius,
is introduced in order to make the potential $H$ dimensionless.
In the case of the \BBz\ decay one can use the closure approximation, replacing
$E_m$ by an appropriate mean value.
(This is justified because we expect that the momentum 
of the virtual neutrino is determined by the
uncertainty relation $q \sim 1/r \sim$ 100 MeV,
thus the variation of $E_m$ from state to state can be neglected.) 
The contributions of the two electrons 
are then added coherently, and thus the neutrino
potential to use is
\be
H(r) = [H_1(r,A_1) + H_2(r,A_2)]/2 \approx H(r,\bar{A}) ~,
\label{eq:hr}
\ee
where $\bar{A} = \bar{E}_m - (M_i + M_f)/2$
and $\bar{E}_m $ is the average energy of the
intermediate nucleus.
 The potential $H(r)$ only very weakly depends
on $m_j$ as long as the neutrino mass is less than $\sim$ 10 MeV.

For the ground state to ground state, i.e., $0_i^+ \rightarrow 0_f^+$
transitions, it is enough to consider $s$-wave outgoing electrons, and   
the nonrelativistic approximation for the nucleons. The nuclear part
of the amplitude then turns into a sum of the Gamow-Teller
and Fermi nuclear matrix elements, where the superscript $0\nu$ is used
to signify the presence of the neutrino potential $H(r)$:
\be
|M_{0\nu}| \equiv M_{GT}^{0\nu} - \frac{g_V^2}{g_A^2} M_F^{0\nu} =
\langle f | \sum_{lk} H(r_{lk},\bar{A}) \tau_l^+ \tau_k^+
\left( \vec{\sigma}_l \cdot \vec{\sigma}_k - \frac{g_V^2}{g_A^2} \right) 
| i \rangle ~.
\label{eq:0nunme}
\ee
The summation is over all nucleons, $| f \rangle ~(| i \rangle)$
are the final (initial) nuclear states, and $g_V (g_A)$ are the
vector (axial vector) coupling constants. Such an expression
is now analogous to the allowed approximation of the ordinary
beta decay.

Thus, in the approximations described above, which are quite accurate,
the transition amplitude for a Majorana neutrino of mass $m_j$
is simply a product of $m_j$ and the above combination of the
nuclear matrix elements. However, since in each of the
two vertices an electron
is emitted, the mixing amplitude $U_{ej}$ apears in each of them,
and the physical \BBz\ reaction amplitude
contains the factor 
$U_{ej}^2$ (not $|U_{ej}|^2$) and is proportional to the factor
\be
\langle m_{\nu} \rangle = \left| \sum_j m_j U_{ej}^2 \right| ~,
\ee
where the sum is  only over light neutrinos with $m_j < $ 10 MeV
(for heavier ones one cannot neglect the mass in the
neutrino propagator, Equation \ref{eq:prop}). The quantity
\mnu\ is the effective neutrino mass.
Since $U_{ej}^2$ and not $|U_{ej}|^2$ appear in \mnu,
its value depends on the Majorana phases discussed above.

To obtain the decay rate,
the reaction amplitude has to be squared,
and multiplied by the corresponding phase space integral,    
which in this case, see Equation \ref{eq:0nurate}, is simply the two-electron
phase space integral proportional to
\be
G^{0\nu} \sim \int F(Z,\epsilon_1) F(Z,\epsilon_2) 
p_1 p_2 \epsilon_1 \epsilon_2
\delta(E_0 -  \epsilon_1 -  \epsilon_2) d\epsilon_1  d\epsilon_2 ~,
\ee
where $E_0$ is the available energy (the sum electron kinetic energy peak
is at $Q = E_0 - 2m_e$). $F(Z,\epsilon)$ is the usual Fermi
function that describes the Coulomb effect on the outgoing electron.

Summarizing, if the \BBz\ decay is mediated by the exchange of a light massive
Majorana neutrino (the assumption that we wish to test), the half-life
is
\be
[ T_{1/2}^{0\nu} (0^+ \rightarrow 0^+)]^{-1}
~=~ G^{0\nu}(E_0,Z) 
\left| M_{GT}^{0\nu} - \frac{g_V^2}{g_A^2} M_F^{0\nu} \right|^2
\langle m_{\nu} \rangle^2 ~,
\label{eq:0nut}
\ee
where $G^{0\nu}$ is the exactly calculable phase space integral,
\mnu\ is the effective neutrino mass and $M_{GT}^{0\nu}$,
$M_F^{0\nu}$ are the nuclear matrix elements, defined in 
Equation \ref{eq:0nunme}.
The way these  nuclear matrix elements are evaluated, and the 
associated uncertainty, is discussed in the next Section.
(As explained earlier, the neutrino mass appears in the amplitude
in the combination $m_j/\omega_j \ll 1$; the denominator $\omega_j$
has been absorbed in the neutrino potential $H(r)$.)

Thus, if an upper limit on  \BBz\ rate is experimentally
established, and the nuclear matrix elements are known,
one can deduce the corresponding upper limit
on \mnu. On the other hand, if \BBz\ is observed, one can deduce
the appropriate value of \mnu. That is a justified procedure, however,
only if the exchange of the light Majorana neutrino, discussed
above, is indeed the mechanism responsible for the decay. There
in no way to decide on the mechanism when only the decay rate is known.
However, a general theorem \cite{SV82} states that once \BBz\ has been observed,
in gauge theories the Majorana neutrino mass necessarily arises.
But the magnitude
of the corresponding neutrino mass is difficult to estimate if the exchange
of a virtual light Majorana neutrino is not the dominant mechanism
of the \BBz\ decay.

\subsection{\BBz\ decay and oscillation parameters}

Let us assume that there are $N$ massive Majorana neutrinos $\nu_i, i = 1,\ldots,N$.
In that case the weak eigenstate neutrinos $\nu_e, \nu_{\mu}$ and $\nu_{\tau}$
can be expressed as superpositions of $\nu_i$ using the $3 \times N$ mixing
matrix $U_{\ell i}$. In particular,
electron neutrinos are
then superpositions,
\be
\nu_e = \sum_i^N U_{e i} \nu_i ~,
\ee
and the rate of the \BBz\ decay is proportional to 
(see Equation \ref{eq:0nut} and Refs. \cite{Doi81,Wolf81})
\be
\langle m_{\nu} \rangle^2 = \left| \sum_i^N  U_{e i}^2 m_i \right|^2 ~ =  
\left| \sum_i^N  |U_{e i}|^2 e^{\alpha_i} m_i \right|^2 ~, ~~({\rm all~~} m_i \ge 0) ~.
\label{eq:mnueff}
\ee
This quantity depends, as indicated, 
on the $N-1$ Majorana phases $\alpha_i/2$ of the matrix $U$
discussed in subsection 2.1 which are irrelevant in neutrino oscillation
experiments that do not change the total lepton number.

If CP is conserved  $\alpha_i = k\pi$, but generally any values
of  $\alpha_i$ are possible. Thus,  \mnu\ could be complex and
cancellations in the sum are possible.
(For example, a Dirac neutrino corresponds to a pair of degenerate
Majorana neutrinos with  $e^{\alpha_i} = \pm 1$ 
whose contribution to \mnu\ exactly cancel.
More generally, some models, such as the Zee model \cite{Zee80},
postulate that \mnu = 0.)

\begin{figure}[htb]
%\epsffile{combined.eps}
\epsfysize=4.0in \epsfbox{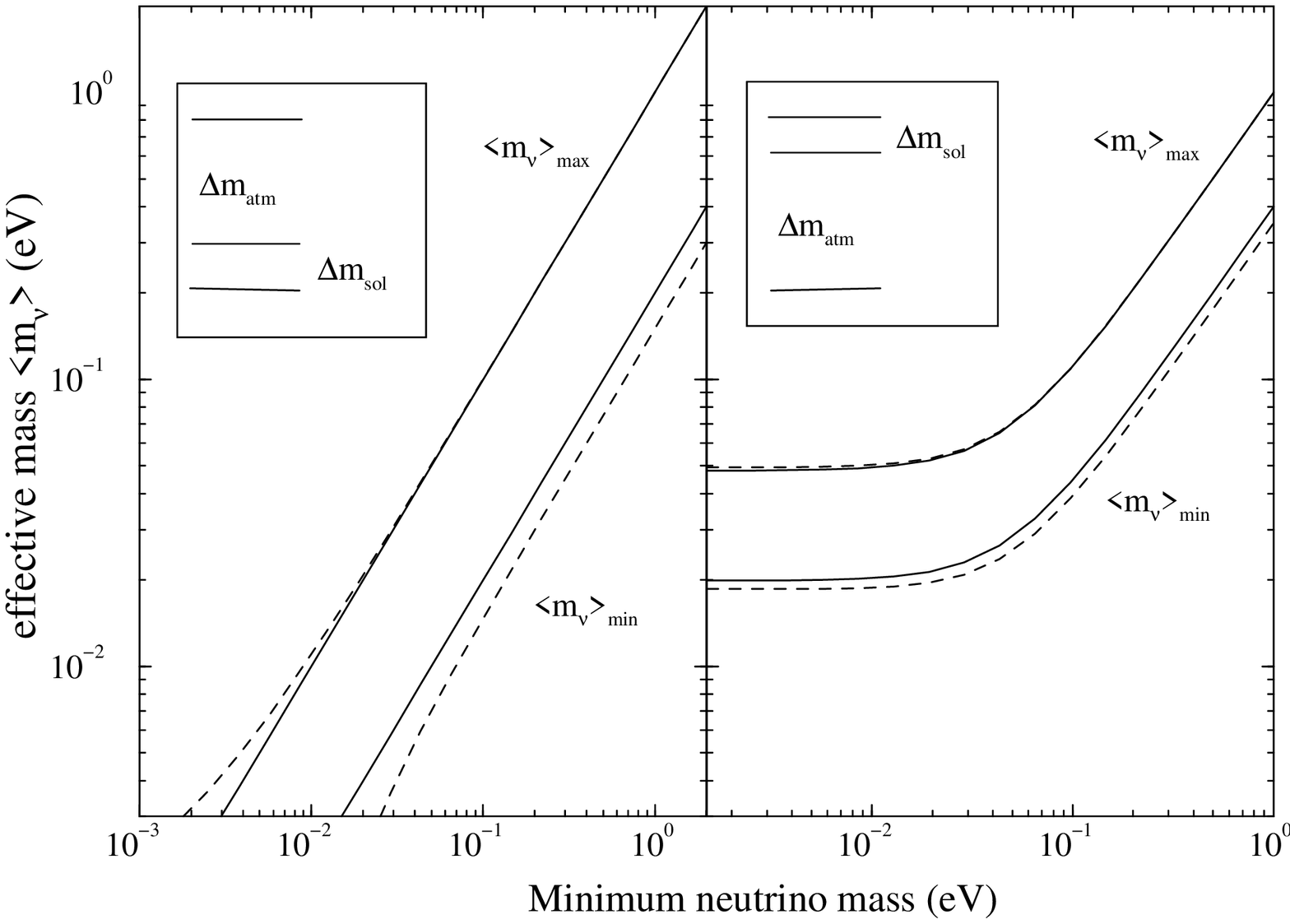}
%\centerline{\epsfbox{combined.eps}}
\caption{Effective mass \mnu\ as a function of the smallest neutrino
mass $m_{min}$. The left panel is for the normal mass hierarchy,
as indicated in the insert (not to scale), and the right panel is for the inverted
hierarchy. Both panels are evaluated for the LMA solar solution
with $\Delta m^2_{atm} = 2.4 \times 10^{-3}$ eV$^2$,  
$\Delta m^2_{sol} = 4.5 \times 10^{-5}$ eV$^2$, and 
$|U_{e2}|^2 = 0.3$. The full lines show \mnu$_{max}$ and
\mnu$_{min}$, defined in Equation \protect{\ref{eq:minmax}},
 for $U_{e3} = 0$ and the dashed lines use the
maximum value $|U_{e3}|^2 = 0.025$ allowed by the \textsc{Chooz} and 
\textsc{Palo Verde} reactor experiments \protect\cite{Chooz99,Palo01}.} 
\label{fig:bb_lma}
\end{figure}

While the quantity \mnu\ depends on the unknown phases $\alpha_i$,
the upper and lower limits of \mnu , 
$\langle m_{\nu} \rangle_{max}$ and  
$\langle m_{\nu} \rangle_{min}$,
depend only on the absolute values of the mixing angles \cite{Viss99},
\be
\langle m_{\nu} \rangle_{max} = \sum_i |U_{ei}|^2 m_i ~, 
~\langle m_{\nu} \rangle_{min} = 
{\rm max}[ (2|U_{ei}|^2m_i -\langle m_{\nu} \rangle_{max}), 0 ]  ~.
\label{eq:minmax}
\ee
Thus, if the search for \BBz\ is successful and the value
of \mnu\ is determined, and at the same time
the mixing angles $|U_{ei}|^2$ and the mass square differences
$\Delta m_{i j}^2$ are known from oscillation experiments,
a {\bf range} of absolute values of the neutrino masses
can be deduced. This is illustrated in Figure \ref{fig:bb_lma}
where we assumed that $N = 3$,
that the Large Mixing Angle (LMA) solution
of the solar neutrinos is correct, and that the
atmospheric neutrino problem requires maximum
mixing of the $\mu$ and $\tau$ neutrinos.  Note that we have to consider
two possibilities, the normal and inverted hierarchies 
(see the inserts in Figure \ref{fig:bb_lma}) because given
the information, we cannot distinguish between them.
(Note that the uncertainty in the mixing parameters is not
included in Figure \ref{fig:bb_lma}.)

Naturally, if another constraint exists, for example a successful
determination of the neutrino mass square $\sum_i |U_{e i}|^2 m_i^2$
in the tritium $\beta$ decay experiments, one can use the knowledge of \mnu\
to determine or constrain the phases $\alpha_i$.

There have been numerous analyses of the existing data that correlate
the current results of the neutrino oscillation searches, \BBz\ experiments,
tritium beta decay experiments, etc. We list here only a subset
of the corresponding papers, and apologize for omission of other
work (see \cite{Viss99,Bil99,KPS01,Mat01,Cza01,PPW01,FSV02}).

Altogether, one cannot predict, in general, what the value of \mnu\
ought to be using the present knowledge. On the other hand, as shown
in Figure \ref{fig:bb_lma} for the currently most likely oscillation
scenario, one can show that certain classes of solutions, such
as the inverted hierarchy, or the normal  hierarchy with
the smallest neutrino mass $\gg \sqrt{\Delta m_{sol}^2}$ 
(degenerate neutrino spectrum) lead to potentially observable
\BBz\ decay.

\subsection{\BBz\ decay and other lepton number violating processes}

The \BBz\ decay is not the only possible observable manifestation of 
lepton number violation. Muon-positron conversion, 
\be
\mu^- + (A,Z) \rightarrow e^+ + (A,Z-2) ~,
\label{eq:muon_conv}
\ee
or rare kaon decays $K_{\mu\mu\pi},~K_{ee\pi}$ and $K_{\mu e \pi}$,
\be
K^+  \rightarrow  \mu^+ \mu^+ \pi^-~, K^+  \rightarrow  e^+ e^+ \pi^-  ~,
K^+  \rightarrow \mu^+ e^+ \pi^- ~,
\label{eq:kaon}
\ee
are examples of processes that violate total lepton number
conservation and where good limits on the corresponding branching
ratios exist. (See Ref. \cite{Zuber00} for a more complete discussion.)

Like the \BBz\ decay, these processes can be mediated by the exchange
of a virtual massive Majorana neutrino. In that case their rate is 
proportional to the quantity analogous to \mnu\ ,
\be
\langle m_{xy} \rangle \equiv \sum_i U_{xi} U_{yi} m_i ~,
\ee
involving other lines of the neutrino mixing matrix $U$
(in \BBz\ the relevant quantity is \mee). 
Note that, again like in  the \BBz, other mechanisms are
possible and might lead to faster rates.

Study of such
decays, in principle, would allow one to constrain, or determine, the otherwise
inaccessible Majorana phases in $U$. However, the present and foreseeable
future of the experimental search has not reached the required sensitivity.

Considerable experimental effort has been devoted to the study  of the
$\mu^- \rightarrow e^+$ conversion,
with the best limit \cite{Sindrum98}
\be
\frac{\Gamma ({\rm Ti}  + \mu^-  \rightarrow e^+ + 
{\rm Ca_{gs}})}{\Gamma ({\rm Ti} + \mu^-  \rightarrow \nu_{\mu} + {\rm Sc})} 
< 1.7 \times 10^{-12} ~(90\% ~CL) ~,
\ee
and a substantial improvement is anticipated in proposed experiments.
That branching ratio limit can  be expressed as the limit on 
\be
\langle m_{\mu e} \rangle \equiv
\langle \sum_i U_{e i} U_{\mu i} m_i \rangle < 17 (82) {\rm MeV} ~,
\ee 
where the two limiting values reflect the
dependence on the spin state (0 or 1) of the created proton pair.

Similarly, for the  $K_{\mu\mu\pi}$ decay the branching ratio is presently
restricted to (see Ref. \cite{App00} where the limits on the other lepton 
number violating $K$ decays are described),
\be
\frac{\Gamma (K^+  \rightarrow \pi^- \mu^+ \mu^+ )}{ \Gamma (K^+  \rightarrow {\rm all})}
< 3.0 \times 10^{-9}  ~(90\% ~CL) ~.
\ee 
Following Ref. \cite{LS00}, this branching ratio limit can be expressed as the limit
\be
\langle m_{\mu \mu} \rangle \equiv \sum_i U_{\mu i}^2 m_i ~<~ 4 \times 10^4~ {\rm MeV} ~.    
\ee

Since, obviously, $\langle m_{xy} \rangle \le m_{max}$, the present limits are far
from constraining $U$  for the three light neutrinos whose mass is 
restricted by the tritium $\beta$ decay experiments \cite{t1,t2}
and the positive results of the
atmospheric and solar neutrino oscillation experiments to be $\le \cal{O}$(eV).
The hypothetical subdominantly coupled heavy sterile are also only marginally
constrained \cite{LS00}. 

Nevertheless, it is important to pursue  searches for the 
$|\Delta L| = 2$ processes with  $|\Delta L_{\mu}| \ne 0$,
since they can, generally, yield nonvanishing results
even when \BBz\ decay is vanishing or very slow.

\section{\BBz\ MATRIX ELEMENTS}
%-----------------------------
%
%	last update: 2/17/02
%
%-----------------------------

For all three modes of \BB\ decay (\BBz, \BBt, \BBm) one can separate, 
essentially without loss of accuracy, the
phase space and nuclear parts of the rate formulae. All 
nuclear structure effects are then represented by the nuclear matrix
elements.  In the review \cite{SC98}
one can find a rather complete list of references,
and the results of 
calculation of the nuclear matrix elements.

The half-life for the \BBt\ decay mode can be written in the compact form,
analogous to Equation \ref{eq:0nut} but without the factor \mnu,
\be
[ T_{1/2}^{2\nu} (0^+ \rightarrow 0^+)]^{-1}
~=~ G^{2\nu}(E_0,Z) | M_{GT}^{2\nu} |^2 ~,
\label{eq:2nut}
\ee
where again $G^{2\nu}$ is the exactly calculable phase space integral containing
all the relevant constants, and $M_{GT}^{2\nu}$
is the nuclear matrix element
(there is no Fermi part, due to the isospin conservation),
\be
M_{GT}^{2\nu} = \sum_m ~ \frac{\langle f || \sigma \tau_+ ||m\rangle
\langle m || \sigma \tau_+ || i \rangle }{E_m - (M_i + M_f)/2} ~,
\label{eq:mgt}
\ee
where $|f\rangle$($|i\rangle$)
are the $0^+$ ground states of the
final (initial) even-even nuclei of masses $M_f (M_i)$, 
and $|m\rangle$ are the $1^+$
states in the intermediate odd-odd nucleus of energy $E_m$.
The individual factors in Equation \ref{eq:mgt} have straightforward
physical meaning; the last factor in the numerator
is the amplitude
of the $\beta^-$ decay (or of the forward angle $(p,n)$ reaction)
of the initial nucleus, while the first factor is the
amplitude of the $\beta^+$ decay (or of the $(n,p)$ reaction)
of the final nucleus.  Thus the description of the \BBt\
is equivalent to the description of the full beta strength
functions of both the initial and final nuclei.
The \BBt\ rate is sensitive to details
of nuclear structure, however, because the 
ground state to ground state transition exhausts only a very small
fraction of the double GT sum rule \cite{VEV88}.
Description of the  \BBt\ thus represents a severe test of
the nuclear models used in the evaluation of  $M_{GT}^{2\nu}$.
We list the experimentally determined matrix elements 
for the \BBt\ decay in Table \ref{tab:2nu}. Note that
the nuclear structure effects cause variations by a factor
$\sim 10$ in the matrix elements, i.e. by a factor
$\sim 100$ in the half-lives.

\begin{table}[htb]
\caption{\protect  Summary of experimentally measured \BBt\ half-lives and matrix elements
($^{136}$Xe is an important exception where a limit is quoted).
See text for a discussion of the averaging procedure. In the Isotope column, the symbol
 $^\dagger$ is used for inconsistent results in which case
the uncertainty reflects the spread in the measured values.  In the references column, the $^\dagger$ is used
to indicate the outliers which were not used in the averaging. The nuclear matrix elements were deduced
using the phase-space factors of Ref. \cite{BV92} using the mean  T$_{1/2}^{2\nu}$.}
\label{tab:2nu}
\begin{center}
\begin{tabular}{lrlc}  \hline\hline
Isotope                & T$_{1/2}^{2\nu}$ (y)                   & References                                                      & $M_{GT}^{2\nu}$ (MeV$^{-1}$) \\ \hline
$^{48}$Ca              &        $(4.2 \pm 1.2)\times 10^{19}$          &  \cite{BAL96,BRU00}                                             & 0.05                \\
$^{76}$Ge              &        $(1.3 \pm 0.1)\times 10^{21}$          &  \cite{KLA01a,AVI91,AAL96}                                      & 0.15           \\
$^{82}$Se              &        $(9.2 \pm 1.0) \times 10^{19}$         &  \cite{ELL92,ARN98}                                             & 0.10            \\
$^{96}$Zr$^{\dagger}$  & $(1.4^{+3.5}_{-0.5})\times 10^{19}$    &  \cite{ARN99,KAW93,Wieser01}                                    & 0.12    \\
$^{100}$Mo             & $(8.0 \pm 0.6)\times 10^{18}$          & \cite{DAS95,EJI91a,EJI91c,DES97,ALS97,ASH01},\cite{VAS90}$^{\dagger}$ & 0.22   \\
$^{116}$Cd             & $(3.2 \pm 0.3)\times 10^{19}$          &  \cite{ARN96,DAN00,EJI95}                                       &  0.12           \\
$^{128}$Te$^{(1)}$     & $(7.2\pm 0.3)\times 10^{24}$           & \cite{BER93,CRU93}                                              & 0.025 \\
$^{130}$Te$^{(2)}$     & $(2.7\pm 0.1)\times 10^{21}$           & \cite{BER93}                                                    & 0.017\\
$^{136}$Xe             & $>8.1\times 10^{20}$ (90\% CL)         & \cite{GAV00}                                                    & $<$0.03 \\
$^{150}$Nd$^{\dagger}$ & $7.0_{-0.3}^{+11.8} \times$ 10$^{18}$  & \cite{DES97,ART95}                                              & 0.07 \\
$^{238}$U$^{(3)}$      & $(2.0\pm 0.6)\times 10^{21}$           & \cite{TUR91}                                                    &  0.05 \\ \hline
\end{tabular}
\end{center}

$^{(1)}$deduced from the geochemically determined half-life ratio $^{128}$Te/$^{130}$Te \\
$^{(2)}$geochemical result includes all decay modes; other geochemical determinations only marginally agree \\
$^{(3)}$radiochemical result, again for all decay modes
\end{table}

The nuclear matrix elements defined in Equation \ref{eq:0nunme}
govern both \BBz\ and \BBm\ decay modes. However, the
half-life of the \BBm\ mode depends on the effective
majoron-neutrino coupling constant $\langle g_{\nu \chi} \rangle$,
instead of \mnu\
\be
[ T_{1/2}^{0\nu, \chi} (0^+ \rightarrow 0^+)]^{-1}
~=~ G^{0\nu, \chi}(E_0,Z) 
\left| M_{GT}^{0\nu} - \frac{g_V^2}{g_A^2} M_F^{0\nu} \right|^2
\langle g_{\nu \chi} \rangle^2 ~,
\label{eq:majt}
\ee
where again $ G^{0\nu, \chi}(E_0,Z) $ is the phase-space integral,
tabulated e.g. in Ref. \cite{DKT85}.

Throughout, we discuss only the \BBz\ decay 
mediated by the exchange of a light massive
Majorana neutrino and governed by the nuclear matrix elements (\ref{eq:0nunme}).
The rate of  \BBz\ mediated by other mechanisms, e.g. involving the
right-handed current weak interactions, or the exchange
of only heavy particles (heavy neutrinos, SUSY particles, etc.)
depends on other nuclear matrix elements. Detailed discussion
of their evaluation is beyond the scope of this review,
but an interested reader can find it e.g. in Refs. \cite{SC98,FS98,Pas99}.

There are two basic approaches to the evaluation of the nuclear
matrix elements for both the \BBt\ and \BBz\ decays, the
quasiparticle random phase approximation (QRPA), and the
nuclear shell model (NSM). As pointed out above, the factors entering
the \BBt\ matrix elements are related to other nuclear phenomena,
and thus testable. 
This is not so, or at least it is much more difficult,
for the matrix elements of the \BBz\ mode. It is therefore
less clear how to reliably estimate   the uncertainty involved in their
evaluation. 

QRPA has been the most
popular theoretical tool in the recent past
since it was able to explain the suppression,
relative to the sum rule \cite{VEV88}, of $M_{GT}^{2\nu}$,
and it is easy to use. 
The main physics ingredients
of the method, relevant particularly for the testable \BBt, are the
repulsive particle-hole spin-isospin interaction and the attractive
particle-particle interaction, which clearly play a decisive role in the
concentration of the $\beta^-$ strength in the giant GT resonance,
and the relative suppression of the $\beta^+$ strength and its
concentration at low excitation energies. Together, these two
ingredients are able to explain the values of $M_{GT}^{2\nu}$
provided the empirical parameter $g_{pp}$, the strength
of the particle-particle interaction is adjusted (but its
adjusted value is, reassuringly, near its expected value).
Yet, the QRPA is often criticized for two ``undesirable'' features.
One is the  sensitivity of  $M_{GT}^{2\nu}$ to the $g_{pp}$ value which
decreases the predictive power of the method. The other is the
fact that for a realistic value of  $g_{pp}$ the QRPA solutions are close
to their critical value (so called collapse, beyond which
the solutions of the QRPA equations do not exist). The
collapse indicates a phase
transition, i.e., a rearrangement of the nuclear ground state. QRPA
is meant to describe small deviations from the unperturbed ground state,
and thus is not fully applicable near the point of collapse.
Numerous approaches have been made to extend the range of validity of QRPA,
usually involving corrections to the quasi-boson approximation.
(See, e.g., Ref. \cite{SC98}). 

While extensive work has been done on this
aspect of QRPA, the unresolved question is what effect
on the \BB\ nuclear matrix elements, and in particular on the \BBz\ ones, 
have the more
complicated configurations that are not included in the QRPA
(i.e., the two-quasiparticle states and their iterations).
There is also a lack of detailed nuclear spectroscopy
predictions (beyond the $\beta$ strength function).

Ideally, the nuclear shell model, NSM, is the method of choice for the
evaluation of \BB\ nuclear matrix elements. In it, one chooses
a set (limited basically by the capability of the
present-day computers) of valence single
particle states. Then one finds an effective hamiltonian,
based usually on the free nucleon-nucleon interaction, but modified to describe the
effective nucleon interaction for that particular set. All configurations (or at least
a convergent set of them) are used in the diagonalization of the hamiltonian,
and in the evaluation of the \BB\ nuclear matrix elements. The method is tested,
and the hamiltonian is adjusted,
by requiring that it describes the spectroscopy (level energies and transition
probabilities) of the relevant nuclei. 

Despite the tremendous advances
of the computational techniques, only a limited set of single-particle states
can be included in the NSM.
The effects of single-particle states that are excluded is 
usually simulated in the NSM by using
effective operators, or more simply, by using effective charges.
Unfortunately, it is not clear what effective charges, if any
(or indeed what effective operators), should be used in the NSM evaluation
of the \BBz\ matrix elements $M_F^{0\nu}$ and  $M_{GT}^{0\nu}$,
and it is not clear what related phenomena one can use
to determine them.

Thus, each approach, QRPA and NSM, has its strengths and weaknesses,
and naturally its critics and defenders. Given this situation, it is
customary, although not really justified, to consider the spread
of the theoretically calculated \BBz\ nuclear matrix elements
as a measure of their uncertainty. Clearly, a breakthrough in
the evaluation of these matrix elements, or at least in the estimate
of their uncertainty, would be very welcome. Given the experimental
effort described in the following sections and the importance
of the problem, we hope that a comparable theoretical effort
will emerge and result in radical improvement in the
nuclear matrix element evaluation. 

To better appreciate the difficulty in evaluating the \BBz\
nuclear matrix elements, let us stress once more that the
presence of the neutrino propagator leads to the appearance
 of the ``neutrino potential'' $H(r,\bar{A})$, Equation \ref{eq:hr}.
Then,
\be
M_{GT}^{0\nu} = \langle f | \sum_{lk} \vec{\sigma}_l \cdot \vec{\sigma}_k
\tau_l^+ \tau_k^+ H(r_{lk},\bar{A} )| i \rangle ~~,
\label{eq:mgt0nu}
\ee
\be
M_F^{0\nu} = \langle f | \sum_{lk} 
\tau_l^+ \tau_k^+ H(r_{lk},\bar{A} ) | i \rangle ~~.
\label{eq:mf0nu}
\ee
Here the $l,k$ summation is over all pairs of neutrons (or protons). 
Note that, due to the presence
of $H(r,\bar{A})$, the Fermi matrix element $M_F^{0\nu}$ is nonvanishing 
even if isospin is conserved.

One can now expand the potential  $H(r,\bar{A} )$ in multipoles
corresponding to the various angular momenta of the intermediate
odd-odd nucleus. One finds then, as expected due to the high
excitation energy (or high value of the momentum $q$
of the virtual neutrino), that many multipoles give comparable contributions.
Moreover, the $1^+$ multipole, which is the only one contributing
to the \BBt, is suppressed and contributes very little.
Thus, a correct reproduction of measured
$ M_{GT}^{2\nu}$ is a necessary but insufficient
condition for equally successful evaluation of the
$M_{GT}^{0\nu}$ and $M_F^{0\nu}$ nuclear matrix elements.

It has been often argued that, unlike $ M_{GT}^{2\nu}$,
there is no suppression in $M_{GT}^{0\nu}$ and $M_F^{0\nu}$,
and hence their values are less sensitive to nuclear structure
details. This argument is based on the multipole decomposition,
discussed above, in the various angular momenta and parities
of the virtual states in the intermediate odd-odd nucleus.
It turns out that the contributions of most of the mutipoles
have the same sign, and hence do not interfere with each other.

However, it is possible to expand the corresponding
expression in an equivalent representation in terms of
the angular momenta and parities of the pair of neutrons
that are transformed into the pairs of protons. In this,
equally valid representation, the 
dominant contribution of the $J = 0^+$ pairs
is to a large extent cancelled by the contribution
of all other,  $J \ne 0^+$, pairs which have an
opposite sign.   
Thus, also in the \BBz\ case the nuclear matrix
elements depend on the small, and presumably poorly determined, pieces
of the nuclear wave function.

Despite our reservations expressed above, in Table \ref{tab:0nu_me} we compare the
\BBz\ half-lives evaluated for \mnu\ = 50 meV with nuclear matrix elements
evaluated in the quoted references chosen to represent
the vast literature on the subject. The spread of the calculated values for the
given parent nucleus gives some indication
of the role played by the nuclear matrix elements.
On the other hand, the spread of half-lives along the columns
in   Table \ref{tab:0nu_me} reflects both effects of the phase space,
and the nuclear matrix elements.  
The methods used to evaluate them are: truncated shell model
\cite{HS84}; nuclear shell model \cite{Caurier99}; QRPA with 
the schematic $\delta$ force interaction 
($\alpha_1'$ = 390 MeV fm$^3$, recalculated for $g_A$ = 1.25) \cite{EVZ88};  
QRPA with $G$-matrix based interaction \cite{Staudt90}; 
renormalized QRPA \cite{FS98,TS95}; and
QRPA without the $p-n$ pairing \cite{Pantis96}.
Clearly, the calculated half-life uncertainty of about
an order of magnitude,
corresponding to a factor of $\sim 3$ in \mnu, can be seen.

\begin{table}
\begin{center}
\caption{\BBz\ half-lives in units of $10^{26}$ years corresponding to
\mnu\ = 50 meV for nuclear matrix elements evaluated in the indicated
references.}
\label{tab:0nu_me}
\vspace{0.3cm}
\begin{tabular}{lcccccl}%
\hline \hline
Nucleus & Ref.: \cite{HS84} & \cite{Caurier99} & \cite{EVZ88} & \cite{Staudt90}
& \cite{FS98,TS95} & \cite{Pantis96} \\
\colrule
$^{48}$Ca & 12.7    & 35.3   &   -   &   -  &    -  & 10.0    \\
$^{76}$Ge & 6.8 & 70.8 & 56.0 & 9.3 & 12.8 & 14.4 \\
$^{82}$Se & 2.3 & 9.6 & 22.4 & 2.4 & 3.2 & 6.0 \\
$^{100}$Mo & - & - & 4.0 &  5.1 & 1.2  & 15.6 \\
$^{116}$Cd &  -   &   -   &  -   &  1.9   & 3.1     & 18.8    \\
$^{130}$Te & 0.6 & 23.2 & 2.8 & 2.0 & 3.6& 3.4 \\
$^{136}$Xe & - & 48.4 & 13.2 & 8.8 & 21.2 & 7.2 \\
$^{150}$Nd${^a)}$ &  -   &  -    &   -   & 0.1    &  0.2    & -    \\
$^{160}$Gd${^a)}$  & -    &  -    &   -   & 3.4    &   -   &  -   \\
\botrule
\end{tabular}
\end{center}

$^{a)}$ deformed nucleus; deformation not taken into account
\end{table}

\section{\BBz\ EXPERIMENTAL OVERVIEW AND PAST \BBz\ EXPERIMENTS}
%-----------------------------
%
%	last update: 2/26/02
%
%-----------------------------

The various detection schemes for \BB\ are effectively outlined 
in Ref. \cite{MV94} so we will only
mention the salient points here. Over the past 15
years, the success of background reduction has resulted 
in a large number of \BBt\ half-life measurements. With
this success behind us, the community is currently focusing 
on the more exciting goal of \BBz. The 
various modes of \BB\ are separated by the differences in their
electron sum energy spectrum and because \BBz\ is 
identified by its distinguishing sum-energy peak, 
direct counting experiments with sufficient
energy resolution are the focus of today's researchers. 
In most of these experiments, the source
also serves duty as the detector. 

Geochemical and radiochemical experiments which were a mainstay 
of \BB\ physics through the 1970s and 1980s,
do not discern between the different modes. 
Thus there is little interest in pursuing these 
techniques further. Instead relatively new technologies 
such as bolometers and scintillating
crystals are receiving attention. Tracking and foil-scintillator 
sandwich experiments are also being
pursued although the source and detector are separate. 
Amazingly, the long-time workhorse of
germanium detectors dominates the present \BBz\ decay results 
and comprise some of the most promising
future proposals.

The study of double beta decay is about suppressing backgrounds. 
Therefore, in this section we summarize
the criteria that make a good \BB\
experiment and then discuss the background issues in general. 
Finally, we describe the various past
and current experiments.

\subsection{Experimental Criteria}

For the best sensitivity to \mnu, one must build a detector 
that maximizes the \BBz\ count rate while minimizing
the background. The signal sensitivity is approximately the 
statistical precision (i.e.,  square root) of the background 
determination. Since the number of background counts increases 
linearly with time, the decay rate sensitivity
scales as the square root of time. In turn, the \mnu\ sensitivity 
scales as the square root of the decay rate,
and therefore as the fourth root of the counting time. 
In an experimet with zero background on the other hand, the \mnu\ limit scales
more quickly as the square root of the counting time.
Explicitly, the limit on \mnu\ can be expressed in terms 
of experimental  parameters as \cite{MOE91a} 
\begin{eqnarray}
\langle m_{\nu} \rangle & = & (2.50 \times 10^{-8} \mbox{eV})
\left[ \frac{W}{f x \epsilon G^{0\nu} |M_{0\nu}|^2} \right]^{1/2}
\left[ \frac{b \Delta E}{MT} \right]^{1/4}  
{\rm ~~background~limited~~} \nonumber \\
\langle m_{\nu} \rangle & = & (2.67 \times 10^{-8}  \mbox{eV})
\left [ \frac{W}{f x \epsilon G^{0\nu} |M_{0\nu}|^2} \right]^{1/2}
\times \frac{1}{\sqrt{MT}}  {\rm~~zero~background} 
\label{e:mee}
\end{eqnarray}
\noindent where $W$ is the molecular weight of the source material, 
$f$ is the isotopic abundance, $x$
is the number of \BB\ atoms per molecule, $\epsilon$ is the detector efficiency, 
$b$ is the number of background counts per kg$\cdot$year$\cdot$keV, 
$\Delta$E is the energy window for \BBz\ in
keV, $M$ is the mass of isotope in kg, 
$T$ is the live time of the experiment in years, and $|M_{0\nu}|$
is a shorthand for the \BBz\ matrix elements given by Equation \ref{eq:0nut}. 
Some of the criteria that need consideration when optimizing the design of 
a \BBz\ decay experiment are obvious whereas others are more
subtle. It is clear from Equation \ref{e:mee} that one needs 
a large source mass. To reach the 50 meV region
of interest indicated by the oscillation results, approximately a ton of
isotope will be required. Also obvious is that one needs 
a reliable detector technology that preferably is
easy to operate. Since the experiments are usually conducted 
in underground laboratories far from the experimenter's
home, it is a great convenience if the experiment needs minimal maintenance. 

The search for \BBz\ decay is a search for a peak superimposed on a continuum. 
Therefore good energy resolution is a must.
Not only does it improve the signal to background in the peak search, 
but poor resolution would result in the \BBt\
tail extending up into the peak region to become a background itself. 

Natural radioactivity is present in all materials at some level. Thus the
source and detector must be very low in such impurities. 
Furthermore, the detector must be shielded from the environment
and its associated radioactivity. This shielding must be also radiopure. 
Since the total activity of an
impure material will scale as its volume, it is usually an advantage to 
minimize the detector size. This can be
most readily accomplished by employing a detector that also plays the role of the source.

Cosmogenic activities build up in materials through nuclear 
reactions of cosmic ray muons and their secondary
products, especially neutrons. These can be a significant background contribution
both for the source 
and for the shielding material. 
Some materials have no long-lived isotopes and thus have a built-in
safeguard against cosmogenics. For experiments that must fight this problem, 
fabricating the apparatus underground and storing
materials underground can greatly reduce this background. 
Unlike solid sources, a gaseous or liquid source can be continuously purified of
such impurities.

Choosing an isotope with a large Q-value and matrix element 
improves the \mnu\ sensitivity for a given measured half-life.
But the nuclear theory of some isotopes is better understood than others. 
Since the only feasible experiment sensitive to
\mnu\ is double beta decay and since the half-life for this process depends on both 
parameters (Equation \ref{eq:0nut}),
it is advantageous to use a source isotope for which
there is confidence in the theoretical calculations. Also some nuclei, 
$^{100}$Mo for example, have relatively fast \BBt\
rates with respect to the theoretically anticipated \BBz\ rate 
for a given \mnu. As \BBt\ is a potential background, the source
choice may be important for detectors with modest or marginal energy resolution.

Radiochemical and geochemical experiments operate by detecting 
the daughter of the decay. However, since these
techinques integrate over an exposure time, they can not identify 
the mode of decay. But if an experiment could
identify the daughter in coincidence with a real-time measurement 
of the decay energy, it would have a powerful
tool for rejecting many backgrounds. In this case, only \BBt\ would be a background. 
There are several possibilities for
detecting the daughter. If the decay is to an excited state, 
one might be able to observe $\gamma$ rays 
that identify the daughter.  \BB\ candidates are initially in 0$^+$ states 
and in many cases transitions to excited 2$^+$  or 0$^+$
states are possible. However, the Q-value for these excited-state transitions is much smaller than for the ground-state transition
 and therefore the decay rate for a given \mnu\ value is much less. For the 2$^+$ case, the matrix elements are also much
 smaller due to the forbidden nature of the transition. 
One interesting possibility is that of $^{150}$Nd where
 the excited state is relatively low in energy. 
But this excited state decays via internal conversion  
requiring the detection
of a 30-keV x ray in order to observe the daughter. 
The most enticing situation however, is that of the Xe-Ba system.
The optical detection of the Ba daughter ion might be possible. 
This possibility is discussed in Section 5.

\subsection{Backgrounds}
Any extraneous energy deposit in the detector near the \BBz\ Q-value will limit the sensitivity to \mnu. Thus any radioactive
isotope with a Q-value greater than the \BBz\ endpoint may be a potential background. Since the number 
of radioactive isotopes decreases with increasing Q, it is desirable to select a \BB\ candidate
with as large a Q-value as possible. $\beta$- and $\alpha$-emitting decays are easy to shield and thus are only a problem
 if they occur within the detector or on its surface. Penetrating $\gamma$ rays pose a more
difficult problem. In this subsection, we consider the origin of various sources of background and some techniques for 
their mitigation.

\subsubsection{Natural Activity}
The naturally occurring isotopes of U and Th and their daughters are present as impurities
 in all materials at some level. The half-life of the chain patriarch is comparable to the age of the universe
 but very short compared to the half-life sensitivity of the \BBz\ experiments. Therefore even
 a small quantity of U or Th will create a significant background. In particular, $^{214}$Bi 
and $^{208}$Tl have large Q-values and the decay spectra will overlap the endpoint of almost 
all the \BBz\ candidates. Even tracking experiments have difficulty with these two isotopes due to their
  $\beta$ decays which are promptly followed by internal conversion resulting in a 
two-electron event that mimics \BBz\ or \BBt. Careful selection of materials and purification 
have been successful, if difficult, solutions to these problems. In the past decade, great strides
 have been made in purifying some materials. Of particular note are liquid scintillator
 \cite{BEN98}, electroformed Cu\cite{BRO90}, and CVD nickel\cite{BOG00}. The radioactive chains may or 
may not be in equilibrium depending on the
 sample's history because chemical treatment or purification can disproportionately eliminate
 the daughters. 

Radon gas, either $^{222}$Rn or $^{220}$Rn, is especially intrusive and may infiltrate into a detector's
sensitive region. These parents to $^{214}$Bi
 and $^{208}$Tl are mobile and diffuse through many materials. Their daughters
 tend to be charged and stick to dust or any other electrostatic surfaces. Many experiments 
eliminate Rn from the detector vicinity by purging the volume immediately surrounding it with 
N$_2$ gas that has boiled-off from a liquid nitrogen (LN) supply. Because Rn freezes out at LN 
temperatures, the boil-off gas tends to be very low in Rn especially compared to the laboratory
 air being displaced. Some groups have also installed charcoal Rn scrubbers into the laboratory airstream.

Activities such as $^3$H, $^{14}$C, and $^{40}$K are also naturally present but their Q-values are too low  
to interfere with \BBz\ experiments. However, many \BBz\ experiments double as dark matter 
experiments by studying the low end of their energy spectrum for possible elastic scattering of WIMPS.
These experiments have to consider a larger pool of potential background isotopes.

\subsubsection{Cosmogenic and Induced Activities}
Long-lived radioactive isotopes can be produced through various nuclear reactions. Many of these 
isotopes can have decay energies that exceed the \BBz\ Q-value and thus can create a background. 
Such activities can be produced in the detector or the shielding material. In particular, long-lived
 isotopes of the source element can be very difficult to remove by purification. The troublesome 
$^{68}$Ge 271-day activity that builds up in Ge detectors is a good example of this problem. 
Short-lived activities can also create background if created {\it in situ} while the experiment is 
operating. There are several nuclear processes that should be considered and we discuss them in this
section. The magnitude of the background due 
to each process is material dependent, and the flux of the projectiles that induce the activities
 depends on the environment. In fact, many processes are greatly minimized by going 
underground where the cosmic ray flux is decreased.

Neutron capture produces $\gamma$ rays and frequently a radioactive isotope. Since neutrons are neutral
 and difficult to identify with anti-coincidence detectors, they can be a significant problem.
On the Earth's surface most neutrons arise from the
cosmic-ray hadronic component.
 In shallow laboratories, secondary neutrons from cosmic-ray muon interactions can form a large contribution to the total
 neutron flux. In deep sites, however, the neutron flux is dominated by ($\alpha$, n) reactions and
 fission neutrons from the laboratory's rock walls. Siting deep underground, covering the walls with 
shielding material to reduce the overall flux 
inside the laboratory, and placing neutron shielding around the detector can help control this background.

Fast neutron reactions also need consideration. $^{68}$Ge for example is produced 
by fast neutron ($\>$25 MeV) interactions on stable Ge isotopes. 
Above ground the dominant source of these fast neutrons are secondaries 
produced by cosmic rays. Once 
the material is taken deep underground, 
the problem is mitigated for the most part, as only
the residual surface production remains. 
But to estimate the underground production rate at
a given depths, one requires neutron flux data. There are several 
integral measurements of the neutron flux underground along with estimates 
of the neutron production due to muons. 
However, since the neutron flux falls very quickly with energy, 
it is difficult to deduce the higher energy 
flux from these measurements.  A summary of the past measurements and 
a calculation of the fast neutron 
flux is given in Ref. \cite{WANGY01} and a program to measure 
the flux is being developed \cite{ABD00}.

Muons and muon induced electromagnetic showers can produce background also. Going deep 
underground reduces the flux, and veto systems surrounding the detector are used to eliminate
any prompt activity observed in coincidence. 
However, inelastic $\mu$ scattering and $\mu^-$ capture
produce delayed events inside the detector after the $\mu$ signal. 
If this time delay is too long or if the $\mu$ flux
is too high, anti-coincidence techniques won't suffice. 
Additional depth can remedy this problem.
For high-Z materials $\mu^-$ capture dominates over $\mu$ decay \cite{MAC65} 
and the neutron multiplicity is of order 1,
with an energy spectrum extending to many tens of MeV \cite{KOZ85}. 
Many of the solar neutrino, dark matter, and double beta decay 
experiments have analyzed the possible spallation products that might be produced. 
(See for example Ref.
\cite{OCO88} which gives a brief general discussion of the topic.)

\subsubsection{Artificially produced activity}
Artificial radioactive isotopes can also be present in materials. 
For example $\approx$ 10$^{15}$ Bq of $^{239,240}$Pu from 
the above-ground testing of nuclear weapons coats the surface of the Earth.
Nuclear accidents such as that at Chernobyl have introduced long-lived isotopes 
e.g., $^{137}$Cs, $^{90}$Sr, and Pu into the environment also. When considering what 
backgrounds might be present, it is thus prudent to consider 
these exotic possibilities. The noble gas
radioactive isotopes $^{42}$Ar and $^{85}$Kr arise from 
the venting of reactors and atmospheric testing; these
must be considered by Xe \BB\ experiments.

\subsubsection{\BBt\ as a background}
	When searching for the \BBz\ peak, one must consider \BBt\
decay as a potential ultimate background. Near the endpoint energy (Q),
 the \BBt\ spectrum has very little strength.
 But since \Tt\ is much shorter than \Tz, the effect of resolution (shown in Fig. \ref{fig:spect})
 must be considered.  
Roughly speaking, the \BBt\ counts within one peak width ($\Delta$E) 
centered on Q will contribute to the \BBz\ peak region and be a background.
The fraction (F) of the \BBt\ counts in the peak region can be approximated by
\be
F  =  \frac{7Q\delta^6}{m_e}\mbox{.} \\
\ee
\noindent where $\delta$ (= $\Delta E/Q$)
is the FWHM energy resolution expressed as a fraction and $m_e$ is the electron mass. The coefficient
$7$ is for a resolution of 5\%. This coefficient depends moderately on resolution and is 8.5 (5) at
1\% (10\%).
An expression for the \BBz\ signal (S) to \BBt\ background (B) ratio can then be written
\be
\frac{S}{B}  =  \frac{m_e}{7Q\delta^6}\frac{\Gamma_{0\nu}}{\Gamma_{2\nu}} 
 =  \frac{m_e }{7Q\delta^6}\frac{T^{2\nu}_{1/2} }{ T^{0\nu}_{1/2} }\mbox{.}  \\
\ee
\noindent Although this approximation cannot replace a Monte Carlo simulation of an experiment's performance, 
it clearly indicates that good energy resolution is critical. But, in addition, the ratio of \Tz\ (for \mnu\ = 1 eV) 
to \Tt\ can vary from 5,000 to 100,000 depending on the isotope's Q-value and matrix element estimate. Therefore,
the choice of isotope  needs consideration. 

For a S/B of 1, the \mnu\ sensitivity limit due to the \BBt\ background can be estimated as
\be
\langle m_{\nu} \rangle^2  \sim  
\frac{7Q \delta^6}{m_e} \frac{G_{2\nu}}{G_{0\nu}} \frac{|M_{2\nu}|^2}{|M_{0\nu}|^2}\mbox{.}     \\
\ee
\noindent Note that using an asymmetric 0$\nu$ window defined by $Q < E < Q+\Delta E/2$ reduces S by a factor of 2 but
decreases B by a factor of $\approx$16. This is exploited by some experiments where the resolution
is not ideal. A previous calculation of the \BBt\ contribution to 
the upper half of the \BBz\ window differs from this
estimate. The author of Ref. \cite{MOE91} agrees with the present result.

\subsection{The Past Experiments}

We have written this section 
to embellish the work of Ref. \cite{MV94}, not replace it. Thus we have de-emphasized material
already covered there. There has been impressive progress in systematically cataloging \BBt\ rates
as detailed in Ref. \cite{TZ02} and summarized in Table \ref{tab:2nu}. 
In this summary table, we have listed average values of the 10 measured \BBt\ 
half-lives and deduced \Mt\ values from them.
The quoted value for each parent nucleus is the weighted average of the chosen measurements.
We included in the average selected measurements with quoted uncertainties
 small enough to significantly affect the average. To assign the individual uncertainty 
associated with each measurement, we first separately averaged 
the asymmetric statistical and systematic 
errors, and then added the two classes of errors in quadrature. 
The summarized \Tt\ are quoted at the 68\%
confidence level. In the individual measurements where 
the uncertainty range was quoted at a different confidence
level, we scaled the uncertainties so they would correspond to a 68\% confidence range. 
In the case of $^{96}$Zr the measurements
are inconsistent and we chose the spread of the measurements 
as an indication of the uncertainty in 
the measured \Tt. In the case of $^{100}$Mo, 
one measurement was very different than the others and 
we did not use it in the average. The nuclear matrix elements were deduced
using the phase-space factors of Ref. \cite{BV92}.  
This procedure is somewhat arbritrary but the details on the half-life 
measurements can be found in the quoted references.
Although $^{136}$Xe has not had its \Tt\ measured yet, 
it is an isotope under consideration for some of 
the future big experiments and therefore its \Tt\ is of importance. 
We list its limit in the table. 

It is worthwhile to note that in one case, for the $\beta\beta$ decay of $^{100}$Mo,
the transition to the excited $0^+$ state at 1.13 MeV in  $^{100}$Ru has been observed
\cite{Bar99,DeB01}.
That state deexites by emission of two $\gamma$ rays; their observation serves as 
a convenient and clean signature of the decay.  The resulting averaged half-life,
$T_{1/2} = (6.8 \pm 1.2) \times  10^{20}$ y, corresponds to a matrix element of similar
magnitude as the $2\nu$ transition to the $^{100}$Ru ground state, in agreement with
expectations \cite{GV92}. Similar transitions are possible in other $\beta\beta$ decay
candidates, but have not yet been observed.

The \BBz\ half-life limits
have also improved for many isotopes and they are summarized in Table \ref{tab:0nu}. The spread
of calculated \Tz\ given in Table \ref{tab:0nu_me} gives an indication of the uncertainty
in \mnu\ due to the uncertainty in the nuclear physics. 
As mentioned earlier the spread in the half-lives
is about an order of magnitude and thus the spread in deduced \mnu\ would be about a factor of 3.
The best limits come from the $^{76}$Ge experiments and they indicate that \mnu\ $<$ 0.3 - 1 eV.

The half-life limits for the Majoron mode are also improving. 
Like the \BBt\ mode, the \BBm\ spectrum is also
a continuum and therefore the limits, summarized in Table \ref{tab:maj},
are more comparable to \BBt\ than to \BBz. 
We have only considered Majoron decay modes that emit a single Majoron,
and list in Table \ref{tab:maj} the half-life limits and the corresponding
\gnu values.
(See Equation \ref{eq:majt} for the relationship between
\Tc\ and \gnu.) Note that, naturally, the deduced \gnu\ limits depend
on the nuclear matrix elements.  

Interestingly, the best
constraint on \gnu\ comes from $^{128}$Te which has the longest measured total half-life. 
Furthermore, its Q-value is very low, resulting in a relative enhancement of
the  phase space factors for the \BBm\ and \BBz\ modes compared to the \BBt\ mode.
Therefore, even though the observed rate is most likely due to 
the \BBt decay, a conservative 
assumption is to assign all the rate to an exotic mode when estimating parameter 
limits. In the case of 
$^{128}$Te, the limit on \gnu\ of $\leq 10^{-5}$ is the best, 
and for \BBz\ the limit of $\approx 1$ eV is competitive.

In the remainder of this section we discuss
selected recent \BB\ experiments 
that can be considered effective prototypes for future programs.
We have ordered the discussion in the current section to parallel Section 5 according to
experimental technique.

\subsubsection{MIBETA}

The MIBETA experiment \cite{ALE00} used TeO$_2$ crystals as bolometers. These detectors 
exploit the low heat capacity of the crystals at low temperature. A small energy deposit 
therefore results in a significant temperature increase of the crystal. 
The experiment consisted of an array of 20 crystals totaling 6.8 kg. 
Since $^{130}$Te is 33.8\% naturally abundant, enrichment
 was not necessary although two crystals were enriched to 93\% in $^{130}$Te and two others were 
enriched to 95\% in $^{128}$Te. The crystals were arranged into a tower of 5 layers of 4 detectors 
within a dilution refrigerator 3500 mwe underground at the Laboratori Nazionali del Gran Sasso. 
The tower frame was made of OFHC copper with the crystal supports 
composed of Teflon. The temperature sensors were neutron transmutation doped 
germanium thermistors. Old Roman lead ($<$4 mBq/kg $^{210}$Pb) 
was placed inside the cryostat surrounding the tower. 
The dilution refrigerator itself was shielded with
low-activity lead (16 $\pm$ 4 Bq/kg $^{210}$Pb). 
The array was operated at a temperature of $\approx$12
mK with an array-averaged resolution of $\approx$8 keV FWHM (0.3\%)
at the \BBz\ endpoint energy of 2.529 MeV. Since 
thermal detectors are sensitive over their entire volume, 
they are susceptible to surface contamination
 and indeed these crystals did observe a surface $\alpha$ activity 
that contributed significantly to the
\BBz\ window. The cosmogenically produced activities in Te 
are short-lived and therefore posed no significant problem.

\subsubsection{Gotthard tunnel}
The Gotthard Xe experiment\cite{LUE98} used a 5-atm gas time projection chamber with 3.3 kg of
62.5\% enriched $^{136}$Xe. The tracking feature of the detector permitted the identification of
two-electron tracks indicative of $\beta\beta$ decay. The energy resolution at the \BBz\ endpoint
(2.481 MeV) was $\approx 165$ keV FWHM (6.6\%). 
The dominant background for \BBz\ was concluded to be Compton scattered
electrons from natural $\gamma$ activities. These electrons were occasionally misidentified as
two-electron events. Cosmogenic activities are not a serious 
issue for Xe experiments, because there
 are no long lived Xe isotopes and liquid or gaseous Xe can be
continuously purified of non-Xe isotopes.

\subsubsection{Heidelberg-Moscow and IGEX}

The Heidelberg-Moscow\cite{BAU99,KLA01a}  (hereafter referred to as HM) and IGEX\cite{AAL99}  
(International Germanium EXperiment)
collaborations both used Ge detectors 86\% enriched in $^{76}$Ge. HM used 125.5 moles of active 
material whereas IGEX used 90 moles. With comparable masses and run times, the results from the 2
experiments are similar with HM posting a modestly better \Tz\ limit. Since \BBz\ events produce
localized ionization in the detectors and many backgrounds (e.g., Compton scattering of $\gamma$
rays) produce multi-site energy deposits, both experiments 
used pulse-shape discrimination to reduce background.

HM identifies the radioactivities that contribute to the data by their associated peaks within
the spectrum. With these identifications, the response of the detectors 
to was simulated with Monte Carlo. A fit to the actual data using the 
Monte Carlo spectra provided an
indication as to the location of the activity. The conclusion of that work, 
described in Ref. \cite{KLA01a}, is that the copper parts of the cryostat 
contained the majority of the background sources.

IGEX performed a measurement of the cosmogenic activity produced in Ge crystals and
compared that to a calculation of the expected rate based on the measured surface neutron 
flux and neutron interaction  cross sections \cite{AVI92}. The calculated rates and measured 
rates agreed well. It was found that initially $^{68}$Ge was the dominant 
cosmogenic activity with the longer-lived
$^{60}$Co dominating at later times. The
rate of $^{68}$Ge can be determined within each crystal by measuring 
the intensity of the 10.4-keV x-ray peak. 
It was  found that the count rate in the \BBz\ window could be mostly 
attributed to radon intrusion.
However, the authors concluded that with a reduction of radon, Ge activation isotopes would be the 
limiting background source in the IGEX experiment.

The two experiments quote similar background levels in the \BBz\
 region of $\approx$0.20 counts/(keV$\cdot$kg$\cdot$yr) before pulse shape 
discrimination and $\approx$0.06 
counts/(keV$\cdot$kg$\cdot$yr) after. However, remarkably, the two collaborations have
come to very different conclusions as to
the composition of the limiting component of the background in these experiments.
 Resolving this issue is one of the most critical debates 
in experimental \BBz\ research today as the design of 
the next generation of Ge detector experiments depends heavily on its outcome. (See Section 5.1.)

During the final preparations of this manuscript (Jan. 2002), the paper
Ref. \cite{KLA01c} appreared in print. It uses the HM data to claim
evidence for \BBz\ in $^{76}$Ge with a \Tz\ = $(0.8-18.3)\times10^{25}$ y.
 If true, this result would be extremely important and hence
requires extensive substantiation and review. However, by itself, the paper does
not sufficiently support the claim \cite{MPLGroup}. Such
deficiences are not necessarily indications that the claim is
wrong, but they indicate that the assessment of this result
by the \BB\ community will be some time in coming. In particular,
the questions raised in  \cite{MPLGroup} should be answered first by the
authors of Ref. \cite{KLA01c}.

\subsubsection{UCI, ELEGANTS and NEMO}

The University of California at Irvine time projection experiment (TPC) measured
 several isotopes. Each sample was a few tens of grams and placed as a thin foil upon the central
 electrode of a TPC. On either side of this source plane were drift regions for recording a 
three-dimensional image of the ionization trails produced by the \BBt\ electrons. The 
experiment was very successful utilizing the tracking capability to determine several
 kinematic parameters characterizing events. This information was critical to reducing
 backgrounds to a level that allowed the first direct detection of \BB\ \cite{ELL92,ELL87,ELL88}.
 The drawback of the design is the limited amount of source mass as compared to the size and
 complexity of the detector.

The best limits on $^{100}$Mo \BBz\ decay come from the 
ELEGANTS (ELEctron Gamma-ray Neutrino Telescope) 
experiments \cite{EJI96}. The emitted electrons in this experiment traversed drift chambers 
for measuring their trajectories and then passed into plastic scintillator in order to measure 
their energies and arrival times. NaI arrays surrounded the apparatus to provide $\gamma$- and 
x-ray observation. Copper and lead shielding enclosed the detectors. The 171-g $^{100}$Mo source
 was two thin foils situated between the drift chambers. This detector had a diminished \BBz\ 
detection efficiency compared to the Ge detectors or 
bolometers. However, it had additional background rejection power because of its measurement
 of several kinematic parameters. The dominant backgrounds in the \BBz\ window 
were identified to be $^{214}$Bi and $^{208}$Tl contained in the source film and detector elements.

The NEMO-2 experiment (Neutrino Ettore Majorana Observatory) \cite{ARN98,ARN99,ARN96,ARN00,DAS95}
has analyzed the \BBt\ rate for several isotopes. 
The detector had a tracking volume of 1 m$^3$
of He gas with two sides covered by scintillator calorimeters. The tracking volume was bisected by 
a thin source plane and consisted of frames containing crossed Geiger cells. 
An electron was defined 
by a track passing from the source foil to the calorimeter. The three dimensional
track measurements were made using the drift times 
and plasma propagation times of the Geiger cells.
The energy was determined with the calorimeters. 
The various source foils weighed up to about 175 g.

These tracking experiments all had small source masses, 
modest energy resolution, and a complex
apparatus. As a result, it will be a challenge for 
future efforts modeled on these designs to
be competitive in the search for \BBz decay.

\subsubsection{Scintillating crystals}
There has been some progress making large scintillating crystals 
with an appreciable amount
of \BB\ isotope contained. The Bejing group 
used CaF$_2$ \cite{YOU91} to study $^{48}$Ca and placed a lower
limit on the \BBz\ decay rate.
A Kiev-Firenze collaboration\cite{DAN00} has used 
$^{116}$CdWO$_4$ scintillators to measure the \BBt\
half-life of $^{116}$Cd and placed limits on 
the \BBz\ and \BBm\ modes. These experiments are
important in that future experiments are being planned that will exploit similar
crystal technologies.

\begin{table}
\caption{\protect Best reported limits on \Tz. 
The \mnu\ limits and ranges are those deduced by the authors and
their choices of matrix elements within the 
cited experimental papers. All are quoted at the 90\% confidence level 
except as noted. The range of 
matrix elements that relate \Tz\ to \mnu\ can be found in Table \ref{tab:0nu_me}.}
\label{tab:0nu}
\begin{center} 
\renewcommand{\arraystretch}{1.2}
\begin{tabular}{lllc} \\ \hline\hline

Isotope            & T$_{1/2}^{0\nu}$ (y)                       &$\langle m_{\nu}\rangle$ (eV)     & Reference     \\ \hline
$^{48}$Ca          & $>9.5\times 10^{21} (76\%)$                & $<8.3$                           & \cite{YOU91}              \\
$^{76}$Ge          & $>1.9\times 10^{25}$                       & $<0.35$                          & \cite{KLA01a} \\
                   & $>1.6\times 10^{25}$                       & $<0.33-1.35$                     & \cite{AAL99}  \\
$^{82}$Se          & $>2.7\times 10^{22}(68\%)$                 & $<5$                             & \cite{ELL92}  \\
$^{100}$Mo         & $>5.5\times 10^{22}$                       & $<2.1$                           & \cite{EJI96}  \\
$^{116}$Cd         & $>7\times 10^{22}$                         & $<2.6$                           & \cite{DAN00}  \\
$^{128,130}$Te     & $\frac{T_{1/2}(130)}{T_{1/2}(128)} = (3.52\pm 0.11)\times 10^{-4}$  & $<1.1-1.5$ & \cite{BER93} \\
                   &(geochemical)                               &                                  &                 \\
$^{128}$Te         & $>7.7\times 10^{24}$                       & $<1.1-1.5$                       & \cite{BER93} \\
$^{130}$Te         & $>1.4\times 10^{23}$                       & $<1.1-2.6$                       & \cite{ALE00}\\
$^{136}$Xe         & $>4.4\times 10^{23}$                       & $<1.8-5.2$                       & \cite{LUE98} \\ 
$^{150}$Nd         & $>1.2\times 10^{21}$                       & $<3$                             & \cite{DES97} \\ \hline
\end{tabular}

\end{center}
\end{table}

\begin{table}[htb]
\caption{\protect The most restrictive \BBm\ limits. 
The \gnu\ limits are those deduced by the authors of the 
cited experimental papers. All are quoted at the 90\% confidence level 
except as noted. The total geochemical measured decay rate of $^{128}$Te 
is used as its \BBm\ limit. The range of 
matrix elements that relate \Tc\ to \gnu\ can be deduced from the
entries in Table \ref{tab:0nu_me}.}
\label{tab:maj}
\begin{center} 
\renewcommand{\arraystretch}{1.2}
\begin{tabular}{lllc} \\ \hline\hline
Isotope           &  T$_{1/2}^{0\nu,\chi}$ (y)                         &      \gnu\                                      & Reference       \\ \hline
$^{48}$Ca   & $>7.2\times 10^{20}$                                     & $<5.3\times 10^{-4}$                            & \cite{BAR89}     \\
$^{76}$Ge   & $>6.4\times 10^{22}$                                     & $<8.1\times 10^{-5}$                            & \cite{KLA01a}    \\
$^{82}$Se   & $>2.4\times 10^{21}$                                     & $<(2.3-4.3)\times 10^{-4}$                      & \cite{ARN98}     \\
$^{96}$Zr   & $>3.5\times 10^{20}$                                     & $<(2.6-4.9)\times 10^{-4}$                      & \cite{ARN99}     \\
$^{100}$Mo  & $>5.4\times 10^{21}$ (68\%)                              & $<7.3\times 10^{-5}$                            & \cite{EJI96} \\
$^{116}$Cd  & $>3.7\times 10^{21}$                                     & $<1.2\times 10^{-4}$                            & \cite{DAN00} \\
$^{128}$Te  & $>7.7\times 10^{24}$ (geochemical)                       & $<3\times 10^{-5}$                              & \cite{BER93} \\
$^{130}$Te  & $>1.4\times 10^{21}$                                     & $<(2.6-6.7)\times 10^{-4}$                      & \cite{ALE00} \\
$^{136}$Xe  & $>7.2\times 10^{21}$                                     & $<(1.3-3.8)\times 10^{-4}$                      & \cite{LUE98} \\
$^{150}$Nd  & $>2.8\times 10^{20}$                                     & $<1\times 10^{-4}$                              & \cite{DES97} \\ \hline
\end{tabular}

\end{center}
\end{table}

\section{FUTURE \BBz\ EXPERIMENTS AND PROPOSALS}
%-----------------------------
%
%	last update: 2/26/02
%
%-----------------------------

\subsection{The Various Proposals}

It has not been possible to realize an experimental concept 
that features all the criteria described in Section 4.
These criteria are frequently incompatible and thus no past experiment or 
future proposal has been able to optimize 
each simultaneously. We are aware of over 10 ideas or proposals for \BBz\ experiments 
summarized in Table \ref{tab:0nufut}. The fourteen listed proposals 
are divided into 2 groups according to source mass and arranged alphabetically
within each group. Each proposal chooses a different approach to attain this optimization. 
Five of them are substantially developed and have the potential 
to reach the crucial 50 meV region. Although we will briefly mention 
all proposals, we devote particular attention to these five: CUORE, EXO, GENIUS, Majorana, and MOON.

 \subsubsection{CUORE}
The success of the MIBETA experiment has resulted in the CUORE (Cryogenic Underground Observatory for 
Rare Events) proposal \cite{PIR00}. 1000 TeO$_2$ crystals of 750 g each 
would be operated as a collection of
bolometers. The detectors will be collected into 25 separate towers of 40 crystals. Each tower will 
have 10 planes of 4 crystals each. One such plane has already been successfully tested and a single 
tower prototype referred to as CUORICINO has been approved. 

The energy resolution at the \BBz\ peak (2.529 MeV) is expected to be about 5 keV FWHM 
($\approx$ 0.2\%). A low energy threshold of 5-10 keV is anticipated, 
and thus the experiment will also search for dark matter. The 
background has been measured in the first plane to be 
$\approx$0.5 counts/(keV$\cdot$kg$\cdot$y). However a major 
component of this background is due to a surface contamination arising from the use of cerium 
oxide polishing compound which tends to be high in thorium. With this problem solved,
the experimenters project a 
conservative estimate of the background to be $\approx$0.01 counts/(keV$\cdot$kg$\cdot$y).

A major advantage of this proposal is that the natural abundance of $^{130}$Te is 34\% and, thus, 
no enrichment is needed resulting in significant cost savings. As with MIBETA, the cosmogenic
activities within the TeO$_2$ crystals are not a serious concern. On the other hand, the crystal
 mounts and cryostat form a significant amount of material close to the bolometers. Much of the 
cryostat is shielded with Roman period lead but a fair quantity of copper and Teflon remain close 
to the crystals.

\subsubsection{EXO}
The Enriched Xenon Observatory (EXO) \cite{EXO00} 
proposes to use up to 10 t of 60-80\% enriched $^{136}$Xe.
The unique aspect of this proposal is the plan to detect 
the $^{136}$Ba daughter ion correlated with the
decay. If the technique is perfected, it would eliminate 
all background except that associated with \BBt. 
The real-time optical detection of the daughter Ba ion, initially suggested in \cite{MOE91},  might 
be possible if the ion can be localized and probed with lasers. The spectroscopy has been used for 
Ba$^+$ ions in atom traps. However, the additional technology to 
detect single Ba ions in a condensed medium 
or to extract single Ba ions from a condensed medium and trap them must be 
demonstrated for this application. To optically detect the alkali-like Ba$^+$ ion, it is excited 
from a 6$^2$S$_{1/2}$ ground state to a 6$^2$P$_{1/2}$ with a 493-nm laser. Since this excited state
 has a 30\% branching ratio to a 5$^4$D$_{3/2}$ metastable state, the ion is detected by re-exciting
 this metastable state to the 6P state via a 650-nm laser and then observing the resulting decay back to the
 ground state. This procedure can be repeated millions of times per second on a single ion and produce a significant signal.

EXO is presently considering two detector concepts: high-pressure-gas Xe TPC  or Liquid Xe (LXe) 
scintillator. The TPC baseline design consists of two 35 m$^3$ modules at $\approx$20 atm for a 
total of 8.4 t of Xe. The Xe would be contained in a non-structural bag within a pressurized buffer
 gas to constrain the Xe to the active region. The spatial resolution and typical $\beta$ particle
 range (5 cm) will permit the identification of the high-ionization-density points at the terminus of the 
beta tracks, aiding in the separation of two-electron events from one-electron backgrounds
 such as Compton scatters. Upon a trigger of an event near the \BBz\ peak energy, 2.481 MeV,  the
 lasers are directed to the decay point to excite the Ba$^+$ ion. One complication is that the
 \BBz\ of $^{136}$Xe produces a Ba$^{++}$ ion whereas the spectroscopy requires a Ba$^+$ ion. 
Because Xe is a tightly bound atom, charge exchange with the Ba ion is unlikely and
 a quenching gas is required to neutralize one stage of ionization. 

The EXO LXe concept has the advantage of being much smaller than the TPC due to the high density
 of LXe. The scintillation readout has better energy resolution but cannot spatially resolve the
 high ionization points. The higher density makes the scattering of the laser light too great to optically detect
 the Ba$^+$ {\it in-situ}. However, once the Ba ion is localized via its scintillation 
and ionization, it might be extracted 
via a cold finger electrode coated in frozen Xe (M. Vient, unpublished observation, 1991). 
The ion is electrostatically attracted to
 the cold finger which later can be heated to evaporate the Xe and release the Ba ion into a radio 
frequency quadrupole trap. At that point, the Ba$^{++}$ is neutralized to Ba$^+$, laser cooled and
 optically detected. The efficiency of the tagging has yet to be demonstrated and is a focus of current
research.

The collaboration is currently performing experiments to optimize the energy resolution for both 
configurations. The  resolution is a critical parameter as \BBt\ would then be the lone background
 if the Ba tagging is successful. Tests to determine the viability of the Ba
 extraction  process are also being performed. 
The EXO collaboration has received funding to proceed with a 100-kg enriched Xe
detector without Ba tagging. This initial prototype will operate at the Waste Isolation Pilot Plant
(WIPP) in southern New Mexico.

\subsubsection{GENIUS}
The progress and understanding of Ge detectors has been developed over more than 30 years 
of experience. The potential of these detectors lie in their great energy resolution, ease of 
operation, and the extensive body of experience relating to the reduction of backgrounds. 
This potential is not yet exhausted as is evidenced by the GENIUS and Majorana proposals 
that build on the experimenters' previous efforts.

The GENIUS (GErmanium NItrogen Underground Setup) \cite{KLA01b}
 proposal has evolved from the Heidelberg-Moscow (HM) experiment.
 The driving design principle behind this proposed Ge detector array experiment is the 
evidence that the dominant background in the HM experiment was due to radioactivity external
 to the Ge. (The reader should contrast this with the motivation for the design of the 
Majorana proposal described below.) An array of 2.5-kg, p-type Ge crystals would be operated
 ''naked" within a large liquid nitrogen (LN) bath. By using naked crystals, the external 
activity would be moved to outside the LN region. P-type crystals have a dead layer on the
 external surface that reduces their sensitivity to external $\beta$ and $\alpha$ activity.
 Due to its low stopping power, roughly 12 m of LN is required to shield the crystals from 
the ambient $\gamma$-ray flux at the intended experimental site at 
Gran Sasso. By immersion in LN, the optimal operating temperature 
is maintained without a bulky cryostat and a test of the naked operation of a crystal in a 50 l
 dewar has been successful \cite{KLA98}. The results indicate that the performance of the 
detector was comparable to those operated in a conventional vacuum-tight cryostat system. 
Their measurements also indicate very little cross talk between naked detectors and that 
an extended distance ($\approx$6 m) between the FET and the crystal does not degrade the signal.

The proposal anticipates an energy resolution of $\approx 6$ keV FWHM (0.3\%) and a threshold of 11 keV. The value of
 this low threshold is set by x rays from cosmogenic activities. Using 1 t of 86\% enriched 
Ge detectors, the target mass is large enough for dark matter
 studies. In fact a 40-kg $^{nat}$Ge proof-of-principle experiment has been approved for dark
 matter studies.

\subsubsection{Majorana} The Majorana proposal \cite{MAJ01} (named in honor of Etorre Majorana) involves many of the IGEX 
collaborators. Their analysis indicated that $^{68}$Ge contained within the Ge detectors
 was the limiting background for their \BBz\ search. (Contrast this with the GENIUS approach
described above.) The proposal's design therefore 
emphasizes segmentation and pulse shape discrimination to reject this background. The
 electron capture of $^{68}$Ge is not a significant problem but $^{68}$Ge decays
 to the $\beta^+$ emitting $^{68}$Ga. This isotope can create background in the \BBz\ 
window if one of the annihilation $\gamma$ rays converts within the crystal. The energy 
deposits of the positron and $\gamma$ ray may polute the peak window in energy, but the 
deposits will be separated in space. In contrast, a \BBz\ event will have a localized 
energy deposit. Segmentation of the crystals permits a veto of such events. Furthermore,
 distinct ionization events will have a different pulse shape than a localized event. 
Therefore pulse shape analysis can also help reject background. 
Majorana plans to use 210, 86\% enriched, segmented Ge crystals for a total of 500 kg of
 detector. The cryostat would be formed from very pure electroformed Cu ($<$ 25 $\mu$Bq/kg $^{226}$Ra,
9 $\mu$Bq/kg $^{228}$Th)\cite{BRO90}.

\subsubsection{MOON}

The MOON (Mo Observatory Of Neutrinos) proposal \cite{EJI00} plans to use $^{100}$Mo as a \BBz\ source
 and as a target for solar neutrinos. This dual purpose and a sensitivity to low-energy supernova electron
neutrinos \cite{EJI01}
 make it an enticing idea. $^{100}$Mo has a high Q-value (3.034 MeV), which results in a large
phase space factor and places the \BBz\ peak region well above most radioactive backgrounds. It also 
has hints of a favorable
 \Mz\ but unfortunately it has a fast \Tt. The experiment will make energy and
angular correlation studies of \BB\ to select \BBz\ events and to
reject backgrounds. The planned MOON configuration is a supermodule of scintillator and Mo ensembles. One
option is a module of plastic fiber scintillators with thin (0.03 g/cm$^2$)
layers of claded Mo, which are arranged to achieve a position resolution 
comparable to the fiber diameter (2-3 mm). A total of 34 tons of natural Mo would be required. 

As a solar neutrino detector, $^{100}$Mo has a low threshold: 168 keV, and the estimated 
observed event rate is $\approx$160/(ton $^{100}$Mo$cdot$year) without neutrino oscillations. 
It is sobering to realize
 that the primary background for the delayed-coincidence solar neutrino signal
 is accidental coincidences between \BBt\ decays. 

The project needs Mo and scintillator radioactive impurity levels of better than 1 mBq/ton.
 This can be achieved by carbonyl chemistry for Mo and plastics can be produced cleanly. 
However, the total surface area of the Mo-scintillator modules is $\approx$26000 m$^2$.
 Dust, being electrostatically charged, tends to garner Rn daughters and 
becomes radioactive. Keeping these surfaces clean of dust during production and assembly will be a challenge.
 Liquid scintillator and bolometer options that would avoid this large surface area are also being considered. 
The simulations of the design indicate that the energy resolution for the \BBz\ peak will be
 $\approx$7\% which is at the upper end of the range of feasibiltiy for a sub 50 meV \mnu\ experiment.
The bolometer option would also remove the resolution concerns.
Use of enriched  $^{100}$Mo is feasible, and would reduce
the total volume of the detector and source ensemble resulting in a lower
internal radioactivity contribution to the background by an order of magnitude.

\subsubsection{OTHER PROPOSALS}

There are too many proposals for detailed description so we have summarized those
of which we are aware in Table \ref{tab:0nufut} and mention them here. The CAMEO proposal \cite{BEL01}
 would use 1000 kg of scintillating $^{116}$CdWO$_4$
crystals situated within the Borexino apparatus. The Borexino liquid scintillator would provide
shielding from external radioactivity and light piping of crystal events to 
the photomultiplier tube (PMT) array surrounding
the system. Early phases of the program would use the Borexino counting test facility. Similarly, 
the CANDLES proposal\cite{KIS01} (CAlcium floride for study of Neutrino and Dark matter by Low Energy Spectrometer) plans to 
imerse CaF$_2$ in liquid scintillator. The scintillation light from the $\beta\beta$ of $^{48}$Ca will be detected
via PMTs. The low isotopic abundance (0.187\%) of $^{48}$Ca requires a very
large operating mass. Two groups \cite{DAN01,WANGS01} have been studying the use of  GSO crystals (Gd$_2$SiO$_5$:Ce) for the study of
$^{160}$Gd.

COBRA (CdTe O neutrino double Beta Research Apparatus) 
\cite{ZUB01} would use CdTe or CdZnTe semiconductors to search for \BBz\ in either Cd or Te. 1600 1-cm$^3$ 
crystals would provide 10 kg of material. GEM is a proposal \cite{ZDE01} that is very similar to that
of GENIUS. However, much of the LN shielding would be  replaced with high purity
water.

The Drift Chamber Beta-ray Analyzer (DCBA) proposal \cite{ISH00} is for a three-dimensional tracking chamber in a 
uniform magnetic field. A drift chamber inside a solenoid and cosmic-ray veto counters comprises the detector.
Thin plates of Nd would form the source. The series of NEMO experiments is progressing with NEMO-3 \cite{NEMO3}
beginning operation in 2002. In concept the detector is similar to NEMO-2. That is, it 
contains a source foil enclosed between tracking chambers that is itself enclosed within a scintillator
array. NEMO-3 can contain a total of 10 kg of source and plans to operate 
with several different isotopes,
but with $^{100}$Mo being the most massive at 7 kg. The collaboration is also discussing 
the possibility
of building a 100-kg experiment that would be called NEMO-4.

There are two additional groups proposing to use $^{136}$Xe to study \BBz. Caccianiga and Giammarchi \cite{CAC01} propose to 
dissolve 1.56 t of enriched Xe in liquid scintillator. The XMASS \cite{XMASS} collaboration proposes to use
10 t of liquid xenon for solar neutrino studies. The detector would have sensitivity to \BBz.

\subsubsection{ISOTOPE ENRICHMENT}

Only the enrichment facilities of Russia can enrich materials in the 1-ton quantities that
are required for the future proposals. It has several centrifuge enrichment facilities. 
One of these, the Electro Chemical Plant (ECP)
in Krasnoyarsk, can produce $\approx$30 kg/year of enriched $^{76}$Ge material in
apparatus that have not been used for uranium enrichment. With some modest improvements, they
could increase production to $\approx$200 kg/year.  The $^{136}$Xe production rate
at these facilities is estimated to be 2 t/year. With several plants throughout Russia and fairly easy 
expansion, the total capacity is large enough that two
samples of isotope could be produced simultaneously.

\begin{table}[htb]
\caption{\protect Proposed or suggested future \BBz\
experiments, separated into two groups based on the magnitude of the proposed isotope mass.
The \Tz\ sensitivities are those  estimated
by the collaborators but scaled for 5 years of data taking. 
These anticipated limits should be used with caution since
they are based on assumptions about backgrounds for experiments 
that do not yet exist. Since some proposals are more conservative than others in their
background estimates, one should refrain from using this table to contrast
the experiments. The range of 
matrix elements that relate \Tz\ to \mnu\ can be found in Table \ref{tab:0nu_me}.}
\label{tab:0nufut}
\begin{center} 
\renewcommand{\arraystretch}{0.9}
\begin{tabular}{lclc} \\ \hline\hline
                           &            &                                                                 &  Sensitivity to         \\
Experiment                 &	Source     & Detector Description                                            & $T_{1/2}^{0\nu}$ (y)    \\ \hline
COBRA\cite{ZUB01}          &$^{130}$Te  &	10 kg CdTe semiconductors                                       & $1 \times 10^{24}$   \\
DCBA\cite{ISH00}           &$^{150}$Nd  &	20 kg $^{enr}$Nd layers between tracking chambers             & $2 \times 10^{25}$     \\
NEMO 3\cite{NEMO3}         &$^{100}$Mo  &	10 kg of \BBz\ isotopes (7 kg Mo)  with tracking                 & $4 \times 10^{24}$    \\ \hline
CAMEO\cite{BEL01}          &$^{116}$Cd  &	1 t  CdWO$_4$ crystals in liq. scint.                           & $> 10^{26}$   \\
CANDLES\cite{KIS01}        &$^{48}$Ca   &	several tons of CaF$_2$ crystals in liq. scint.                 & $1 \times 10^{26}$      \\
CUORE\cite{AVI01}          &$^{130}$Te  &	750 kg TeO$_2$ bolometers                                       & $2 \times 10^{26}$    \\
EXO\cite{DAN00}            &$^{136}$Xe  &	1 t $^{enr}$Xe TPC (gas or liquid)                              & $8 \times 10^{26}$    \\
GEM\cite{ZDE01}            &$^{76}$Ge   &	1 t $^{enr}$Ge diodes in liq. nitrogen                          & $7 \times 10^{27}$      \\
GENIUS\cite{KLA01b}        &$^{76}$Ge   &	1 t 86\% $^{enr}$Ge diodes in liq. nitrogen                     & $1 \times 10^{28}$    \\
GSO\cite{DAN01,WANGS01}    &$^{160}$Gd  &	2 t Gd$_2$SiO$_5$:Ce crystal scint. in liq. scint.              & $2 \times 10^{26}$   \\
Majorana\cite{MAJ01}       &$^{76}$Ge   &	0.5 t 86\% segmented $^{enr}$Ge diodes                          & $3 \times 10^{27}$   \\  
MOON\cite{EJI00}           &$^{100}$Mo  &	34 t $^{nat}$Mo sheets between plastic scint.                   & $1 \times 10^{27}$   \\   
Xe\cite{CAC01}             &$^{136}$Xe  & 1.56 t of $^{enr}$Xe in liq. scint.                             & $5 \times 10^{26}$    \\
XMASS\cite{XMASS}          &$^{136}$Xe  &  10 t of liq. Xe                                                & $3 \times 10^{26}$    \\\hline

\end{tabular}

\end{center}
\end{table}

%	COBRA - value quoted  read from their figure
% DCBA - My estimate of T1/2 from their mass limit and choice of matrix element. I scaled by sqrt(time) for 5 years
% NEMO 3 - directly quoted from their table for 5 years
% CAMEO - directly quoted from their sensitivity esitmate for 5 years
% CANDLES - They are not very precise in the mass or run time, but wish to go for 10^26 years
% CUORE - They quote 1.1 x 10^26 for 1 year of counting, I scaled by sqrt(time)
% EXO - Direct quote from their sensitivity table
% GEM - They quote a value of 10^28 y for 10 years of counting, I scaled by sqrt(time)
% GENIUS - Gives 2 x10^28 years for 10 years of run time. I scaled by sqrt(time)
% GSO - Direct quote of Danevich paper for 5 years of run time with 2 t
% Majorana - 4 x 10^27 y is given for 10 years of run time, I scaled by sqrt(time)
% Moon - Ejiri calculated that number for me.
% Xe - they give a number for 1 year of running and I scaled by sqrt(time}
% XMASS - Their transparencies directly quote the number

\section{CONCLUSIONS}
%-------------------------------------------------------------
%
%	BBConclusions.tex
% Double Beta Decay Review
%	With Petr Vogel
%	last update: 2/15/02
%
%
%
%-------------------------------------------------------------

We have reviewed the motivation, present status, and future plans
of the search for \BBz\ decay.
Seeing \BBz\ decay would be a remarkable physics result, 
with important consequences for neutrino physics in particular,
and for the hunt for `physics beyond the Standard Model' in
general.

For $N = 3$ neutrino flavors and mass states, the neutrino mass matrix
contains, in general, nine parameters. Three mixing angles, two
$\Delta m^2$ mass differences, and one CP violation phase can be,
in principle, determined in neutrino oscillation experiments.
Remarkable progress has been made lately in determining some of these
parameters, and great effort is being devoted to verify 
the discoveries already made and
refine and extend the search for neutrino oscillations further.

The remaining parameters of the neutrino mass matrix, 
the absolute mass scale and the two
Majorana phases, can be determined or severely constrained
only by the observation of \BBz\ decay, and/or by further progress
in the tritium endpoint neutrino mass experiments.
Thus, the search for the  \BBz\ decay has become one of the
critical issues of particle physics today.

We should note here again that if a non-zero m$_{\nu}$ is observed
in either \BBz\ decay or in tritium
beta decay (or, ideally, in both), the nuclear matrix element  \Mz\ 
issue will become critical.
To dramatize this problem, consider the following possibility. 
Suppose a non-zero \mnu\ of about 100 meV is indicated by 
an upcoming $^{76}$Ge experiment. The anticipated experimental 
uncertainty is approximately $\pm25$ meV, and the matrix element uncertainty 
of a factor of 2-3 would dominate the total uncertainty on 
the mass parameter. Hence the possible range of effective neutrino mass 
would be 33 meV $ < \langle m_{\nu} \rangle <$ 300 meV. 
Keep this range in mind and look at the
right hand side of Fig. \ref{fig:bb_lma}. One sees that this matrix element
uncertainty alone is large enough 
to compromise physics conclusions regarding the mass heirarchy 
at small minimum neutrino masses.  
Therefore, we use this opportunity to express our hope that the 
improved sensitivity of the upcoming \BBz\ experiments will promote
an increased interest in the nuclear theory of double beta decay.

Altogether, the field of  \BB\  decay has seen a
great rebirth of interest stimulated by the recent neutrino physics results. 
The oscillation experiments indicate that neutrinos do have
 mass and, in particular, at least one neutrino has a mass greater than 
$m_{scale} \approx$50 meV. 
The upcoming \BBz\ experiments will have a sensitivity
to \mnu\ values less than this critical mass scale. 
This is a very exciting 
time for \BBz\ research as it is reasonable to hope for a positive
result within the coming decade.

\section{ACKNOWLEDGEMENTS}
We would like to thank the following colleagues for useful 
discussions concerning this manuscript: Frank Avignone III, Laura Baudis, John Beacom, 
Peter Doe, Hiro Ejiri, Giorgio Gratta, Ryuta Hazama, Harry Miley and Michael Moe.
This work was supported by the US Department of Energy under contracts 
DEFG03-97ER41020 and DE-FG03-88ER40397.

\section{LITERATURE CITED}
%-----------------------------
%
%	last update: 2/18/02
%
%-----------------------------

%%% Numbered Literature Cited

\end{document}